\documentclass[twocolumn]{bmcart}

\newcommand{\final}[2]{{#2}}

\usepackage[utf8]{inputenc}
\usepackage{graphicx}
\usepackage[numbers,sort&compress]{natbib}
\usepackage{mathptmx}      %
\usepackage{pgfplots}
\usepackage{booktabs}
\usepackage{pgfplots}
\usepackage{booktabs}
\usepackage{bm}
\usepackage{placeins}
\usepackage{multirow}
\usepackage{lipsum}
\usepackage{lettrine}
\usepackage{pdflscape}
\usepackage{amsmath}
\usepackage{amssymb}
\usepackage{mathtools}
\usepackage{hyperref}
\usepackage{aasmacros}
\usepackage{array}
\usepackage{float}
\usepackage{ragged2e}
\usepackage{ulem}
\usepackage{color}
\usepackage{capt-of}

\newcolumntype{L}[1]{>{\raggedright\let\newline\\\arraybackslash\hspace{0pt}}m{#1}}
\newcolumntype{C}[1]{>{\centering\let\newline\\\arraybackslash\hspace{0pt}}m{#1}}
\newcolumntype{R}[1]{>{\raggedleft\let\newline\\\arraybackslash\hspace{0pt}}m{#1}}
\newcolumntype{P}[1]{>{\centering\arraybackslash}p{#1}}
\makeatletter
\def\cl@chapter{\@elt {theorem}}
\makeatother
\graphicspath{{Figures/}}

\newcommand{\E}{\mathbb{E}}

\DeclareUnicodeCharacter{00A0}{ }
\startlocaldefs

\endlocaldefs
\begin{document}
\begin{frontmatter}
\begin{fmbox}
\dochead{Research}
\title{Cosmological $N$-body simulations: a challenge for scalable generative models}
\author[
   addressref={aff1},
   corref={aff1},
   email={nathanael.perraudin@sdsc.ethz.ch}
]{\inits{NP}\fnm{Nathana\"el} \snm{Perraudin}}
\author[
   addressref={aff3},
   email={ankitsrivastavanit@gmail.com}
]{\inits{AS}\fnm{Ankit} \snm{Srivastava}}
\author[
addressref={aff2},
email={aurelien.lucchi@inf.ethz.ch}
]{\inits{AL}\fnm{Aurelien} \snm{Lucchi}}
\author[
   addressref={aff3},
   email={tomaszk@phys.ethz.ch}
]{\inits{TK}\fnm{Tomasz} \snm{Kacprzak}}
\author[
addressref={aff2},
email={thomas.hofmann@inf.ethz.ch}
]{\inits{TH}\fnm{Thomas} \snm{Hofmann}}
\author[
addressref={aff3},
email={alexandre.refregier@phys.ethz.ch}
]{\inits{AR}\fnm{Alexandre} \snm{R\'{e}fr\'{e}gier}}

\address[id=aff1]{%
  \orgname{Swiss Data Science Center, ETH Zurich},
    \street{Universit\"atstrasse 25},
    \postcode{8006}
  \city{Zurich},
  \cny{Switzerland}
}
\address[id=aff2]{%
  \orgname{Institute for Particle Physics and Astrophysics, ETH Zurich},
  \street{Wolfgang-Pauli-Str. 27},
  \postcode{8093}
  \city{Zurich},
  \cny{Switzerland}
}
\address[id=aff3]{%
    \orgname{Institute for Machine Learning, ETH Zurich},
    \street{Universit\"atstrasse 6},
    \postcode{8006}
    \city{Zurich},
    \cny{Switzerland}
}

\begin{artnotes}
\end{artnotes}
\begin{abstractbox}
\begin{abstract}
\justifying
Deep generative models, such as Generative Adversarial Networks (GANs) or Variational Autoencoders (VAs) have been demonstrated to produce images of high visual quality.
However, the existing hardware on which these models are trained severely limits the size of the images that can be generated. The rapid growth of high dimensional data in many fields of science therefore poses a significant challenge for generative models.
In cosmology, the large-scale, three-dimensional matter distribution, modeled with {\it $N$-body simulations}, plays a crucial role in understanding the evolution of structures in the universe.
As these simulations are computationally very expensive, GANs have recently generated interest as a possible method to emulate these datasets, but they have been, so far, mostly limited to two dimensional data.
In this work, we introduce a new benchmark for the generation of three dimensional $N$-body simulations, in order to stimulate new ideas in the machine learning community and move closer to the practical use of generative models in cosmology.
As a first benchmark result, we propose a scalable GAN approach for training a generator of $N$-body three-dimensional cubes.
Our technique relies on two key building blocks,
(i) splitting the generation of the high-dimensional data into smaller parts, and
(ii) using a multi-scale approach that efficiently captures global image features that might otherwise be lost in the splitting process.
We evaluate the performance of our model for the generation of $N$-body samples using various statistical measures commonly used in cosmology.
Our results show that the proposed model produces samples of high visual quality, although the statistical analysis reveals that capturing rare features in the data poses significant problems for the generative models.
We make the data, quality evaluation routines, and the proposed GAN architecture publicly available at \url{https://github.com/nperraud/3DcosmoGAN}.
\end{abstract}


\end{abstractbox}
\end{fmbox}
\end{frontmatter}

\section{Introduction}
The recent advances in the field of deep learning have initiated a new era for generative models.
Generative Adversarial Networks (GANs)~\cite{GAN} have become a very popular approach by demonstrating their ability to learn complicated representations to produce high-resolution images \citep{Progressive_GAN}.
In the field of cosmology, high-resolution simulations of matter distribution are becoming increasingly important for deepening our understanding of the evolution of the structures in the universe \citep{Springel2005simulations,Sim_2,Kuhlen2012numerical}.
These simulations are made using the {\it $N$-body} technique, which represents the distribution of matter in 3D space by trillions of particles.
They are very slow to run and computationally expensive, as they evolve the positions of particles over cosmic time in small time intervals.
Generative models have been proposed to emulate this type of data, dramatically accelerating the process of obtaining new simulations, after the training is finished \citep{rodriguez2018fast,Mustafa2019cosmogan}.

$N$-body simulations represent the matter in a cosmological volume, typically between 0.1 - 10 Gpc, as a set of particles, typically between 100$^3$ to 2000$^3$.
The initial 3D positions of the particles are typically drawn from a Gaussian random field with a specific power spectrum.
Then, the particles are displaced over time according to the laws of gravity, properties of dark energy, and other physical effects included in the simulations.
During this evolution, the field is becoming increasingly non-Gaussian, and displays characteristic features, such as halos, filaments, sheets, and voids \citep{cosmic_web_1, cosmic_web_4}.

$N$-body simulations that consist only of dark matter effectively solve the Poisson's equation numerically.
This process is computationally expensive, as the forces must be recalculated in short time intervals to retain the precision of the approximation.
This leads to the need for frequent updates of the particle positions.
The speed of these simulations is a large computational bottleneck for cosmological experiments, such as the Dark Energy Survey\footnote{\url{www.darkenergysurvey.org}}, Euclid\footnote{\url{www.euclid-ec.org}}, or LSST\footnote{\url{www.lsst.org}}.

Recently, GANs have been proposed for emulating the matter distributions in two dimensions \citep{rodriguez2018fast,Mustafa2019cosmogan}.
These approaches have been successful in generating data of high visual quality, and almost indistinguishable from the real simulations to experts.
Moreover, several summary statistics often used in cosmology, such as power spectra and density histograms, also revealed good levels of performance.
Some challenges still remain when comparing sets of generated samples. In both works, the properties of sets of generated images did not match exactly; the covariance matrix of power spectra of the generated maps differed by order of 10\% with the real maps.

While these results are encouraging, a significant difficulty remains in scaling these models to generate three-dimensional data, which include several orders of magnitude more pixels for a single data instance.
We address this problem in this work. We present a publicly available dataset of $N$-body cubes, consisting of 30 independent instances.
Due to the fact that the dark matter distribution is homogeneous and isotropic, and that the simulations are made using periodic boundary condition, the data can be easily augmented through shifts, rotations, and flips.
The data is in the form of a list of particles with spatial positions $x$, $y$, $z$.
It can be pixelised into 3D histogram cubes, where the matter distribution is represented in density voxels.
Each voxel contains the count of particles falling into it.
If the resolution of the voxel cube is high enough, the particle- and voxel-based representations should be able to be used interchangeably for most of the applications.
Approaches to generate the matter distribution in the particle-based representation could also be designed; in this work, however, we focus on the voxel-based representation. 
By publishing the $N$-body data and the accompanying codes we aim to encourage the development of large scale generative models capable of handling such data volumes.

We present a benchmark GAN system to generate 3D $N$-body voxel cubes.
Our design of the novel GAN architecture scales to volumes of 256$^3$ voxels.
Our proposed solution relies on two key building blocks.
First, we split the generation of the high-dimensional data into smaller patches.
Instead of assuming that the distribution of each patch is independent of the surrounding context, we model it as a function of the neighboring patches.
Although splitting the generation process into patches provides a scalable solution to generate images of arbitrary size, it also limits the field of view of the generator, reducing its ability to learn global image features.
The second core idea of our method addresses this problem by relying on a multi-scale approach that efficiently captures global dependencies that might otherwise be lost in the splitting process.

Our results constitute a baseline solution to the challenge. While the obtained statistical accuracy is currently insufficient for a real cosmological use case, we achieve two goals: (i) we demonstrate that the project is tractable by GAN architectures, and (ii) we provide a framework for evaluating the performance of new algorithms in the future.

\subsection{Related work}
Generative models that produce novel representative samples from high-dimensional data distributions are increasingly becoming popular in various fields such as image-to-image translation \cite{Image_1}, or image in-painting \cite{in-painting} to name a few. There are many different deep learning approaches to generative models. The most popular ones are Variational Auto-Encoders (VAE) \cite{VAE}, Autoregressive models such as PixelCNN~\cite{Pixel_CNN}, and Generative Adversarial Networks (GAN) \cite{GAN}.
Regarding prior work for generating 3D images or volumes, two main types of architectures -- in particular GANs -- have been proposed. The first type \cite{achlioptas2017learning, fan2017point} generates 3D point clouds with a 1D convolutional architecture by producing a list of 3D point positions. 
This type of models does not scale to cases where billion of points are present in a simulation, posing an important concern given the size of current and future $N$-body simulations. 
The second type of approaches, including \cite{wu2016learning, mosser2017reconstruction}, directly uses 3D convolutions to produce a volume.
Although the computation and memory cost is independent of the number of particles, it scales with the number of voxels of the desired volume, which grows cubically with the resolution. While recursive models such as PixelCNN~\cite{Pixel_CNN} can scale to some extent, they are slow to generate samples, as they build the output image pixel-by-pixel in a sequential manner. We take inspiration from PixelCNN to design a patch-by-patch approach, rather than a pixel-by-pixel approach, which significantly speeds up the generation of new samples.

As mentioned above, splitting the generation process into patches reduces the ability of the generator to learn global image features. Some partial solutions to this problem can already be found in the literature, such as the Laplacian pyramid GAN \cite{LAP_GAN} that provides a mechanism to learn at different scales for high quality sample generation, but this approach is not scalable as the sample image size is still limited.
Similar techniques are used in the problem of super-resolution \cite{ledig2017photo,lai2017deep,wang2018esrgan}.
Recently, progressive growing of GANs \cite{Progressive_GAN} has been proposed to improve the quality of the generated samples and stabilize the training of GANs. The size of the samples produced by the generator is progressively increased by adding layers at the end of the generator and at the beginning of the discriminator.
In the same direction, \cite{brock2018large,lucic2019high} achieved impressive quality in the generation of large images by leveraging better optimization.
Problematically, the limitations of the hardware on which the model is trained occur after a certain increase in size and all of these approaches will eventually fail to offer the scalability we are after.

GANs were proposed for generating matter distributions in 2D.
A generative model for the projected matter distribution, also called a {\it mass map}, was introduced by \citep{Mustafa2019cosmogan}.
Mass maps are cosmological observables, as they are reconstructed by techniques such as, for example, gravitational lensing \citep{Chang2018massmap}.
Mass maps arise through integration of the matter density over the radial dimension with a specific, distance-dependent kernel.
The generative model presented in \citep{Mustafa2019cosmogan} achieved very good agreement with the real data several important non-Gaussian summary statistics: power spectra, density histograms, and Minkowski functionals \cite{Schmalzing1996minkowski}.
The distributions of these summaries between sets of generated and real data also agreed well.
However, the covariance matrix of power spectra within the generated and real sets did not match perfectly, differing by the order of 10\%.

A generative model working on 2D slices from $N$-body simulations was developed by \citep{rodriguez2018fast}.
$N$-body slices have much more complex features, such as filaments and sheets, as they are not averaged out in projection.
Moreover, the dynamic range of pixel values spans several orders of magnitude.
GANs presented by \citep{rodriguez2018fast} also achieved good performance, but only for larger cosmological volumes of 500 Mpc.
Some mismatch in the power spectrum covariance was also observed.

Alternative approaches to emulating cosmological matter distributions using deep learning have been recently been proposed.
Deep Displacement Model~\cite{He2018deepdisplacement} uses a U-shaped neural network that learns how to modify the positions of the particles from initial conditions to a given time in the history of the universe.

Generative models have also been proposed for solving other problems in cosmology, such as generation of galaxies~\citep{galaxy_gen_1}, adding baryonic effects to the dark matter distribution~\citep{Troester2019painting}, recovery of certain features from noisy astrophysical images~\cite{recovery}, deblending galaxy superpositions~\citep{Reiman2019deblending}, improving resolution of matter distributions \citep{Ramanah2019paintinghalos}.

\begin{figure}[t]
    \centering
    \final{
    }{
    \includegraphics[width=0.95\linewidth]{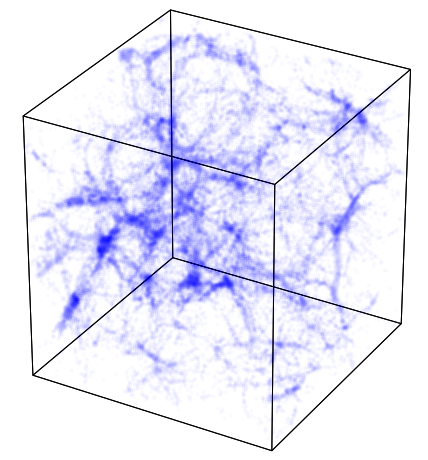}
    }
    \caption{An example $N$-body simulation at current cosmological time (redshift $z=0$).
    The density is represented by particle positions in 3D.
    In this work, the generative models use the representation of this distributon that is based on 3D voxel histogram of the particle positions.
    }
    \label{fig:$N$-body_Simulation}
\end{figure}

\section{The $N$-body data}
\subsection{Cosmological $N$-body simulations}

\label{sec:n_body_simulations}
The distribution of matter, dark matter and other particles in the universe at large scale, under the influence of gravity, forms a convoluted network-like structure called the cosmic web \cite{cosmic_web_1,cosmic_web_2,cosmic_web_3,cosmic_web_4}.
This distribution contains information vital to the study of dark matter, dark energy, and the very laws of gravity \cite{LOG_1,LOG_2,LOG_3}. Simulations of these various computational cosmological models \cite{Sim_1,Sim_2} lead to understanding of the fundamentals of cosmological measurements \cite{Mes_1,Mes_2}, and other properties of the universe \cite{Springel2005simulations}.
These simulations are done using $N$-body techniques.
$N$-body techniques simulate the cosmic web using a set of particles in three dimensional space, and evolve their positions with time.
This evolution is governed by the underlying cosmological model and the laws of gravity.
The end result of an $N$-body simulation is the position of billions of particles in space, as depicted in Figure \ref{fig:$N$-body_Simulation}.
Unfortunately, $N$-body simulations are extremely resource intensive, as they require days, or even weeks of computation to produce them \cite{Nbody_1,Nbody_2}.
Moreover, a large number of these $N$-body simulations is needed to obtain good statistical accuracies, which further increases the computational requirements.

This computational bottleneck opens up a leeway for deep learning and generative models to offer an alternative solution to the problem.
Generative models have the potential to be able to learn the underlying data distribution of the $N$-body simulations using a seed set of $N$-body samples to train on.

There are multiple techniques for running $N$-body simulations, which agree well for large scales, but start to diverge for small scales, around wavenumber $k=1 \ \rm{Mpc}^{-1}$ \citep{schneider2016percent}.
Moreover, baryonic feedback can also affect the small scale matter distribution \citep{Mead2015hmcode, Huang2019baryonic, Barreira2019illustris}, and large uncertainty remains for these scales.

\subsection{Data preprocessing}
We produce samples of the cosmic web using standard $N$-body simulation techniques.
We used L-PICOLA \cite{lpicola} to create 30 independent simulations.
The cosmological model used was $\Lambda$CDM with Hubble constant $\mathrm{H}_0 = 500h = 350 \ \mathrm{km}/\mathrm{s}/\mathrm{Mpc}$\footnote{For cosmological scale, the distance is measured in megaparsec (Mpc) ($1\ Mpc\ = 3.3 \times 10^6\  light \text{-} years$).}, dark energy density $\Omega_\Lambda = 0.724$ and matter density $\Omega_{m} = 0.276$. We used the particle distribution at redshift $z = 0$.
The output of the simulator is a list of particles $1,024^3$ 3D positions.
To obtain the matter distribution, we first convert it to a standard $256^3$ \mbox{3D} cube using histogram binning.
We consider these cubes as the raw data for our challenge and can be downloaded at \url{https://zenodo.org/record/1464832}. The goal is to build a generative model able to produce new 3D histograms.
While $30$ samples might seem as a low number of samples to train a deep neural network, each sample contains a large number of voxels. One can also expand the training data by relying on data augmentation, using various rotations and circular translations as described in Appendix~\ref{sec:augmentation}.
Problematically, the dynamic range of this raw data spans several orders of magnitude and the distribution is skewed towards smaller values, with a very elongated tail towards the larger values.
Empirically, we find that this very skewed distribution makes learning a generative model difficult. Therefore, we first transform the data using a logarithm-based function, as described in Appendix~\ref{sec:input_transform}.
Note that this transform needs to be inverted before the evaluation procedure.

\begin{figure*}[th]
    \centering
    \final{
    }
    {
    \includegraphics[width=1.0\textwidth]{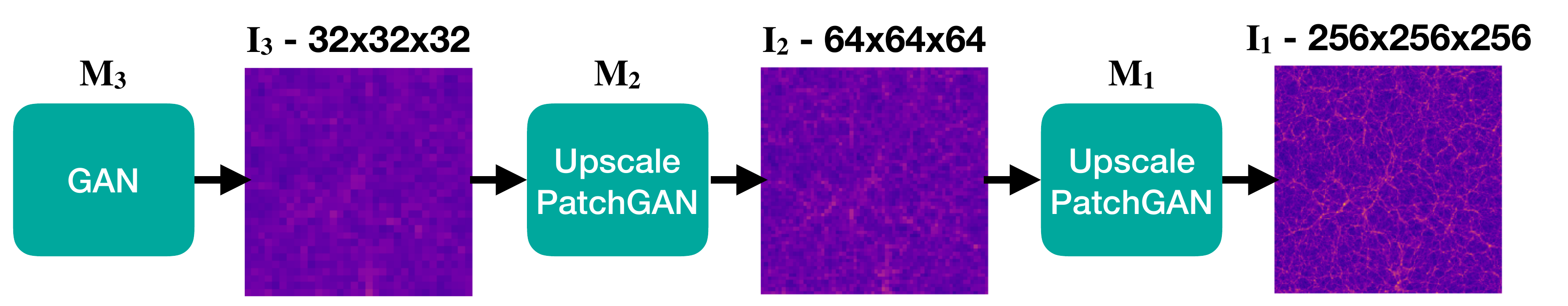}
    }
    \caption{\label{fig:Multi-Scale_Overview} Multiscale approach using multiple intermediary GANs, each learning features at different scales and all trained independently from each other. The same approach can in principle be extended to produce higher-resolution images, potentially adding more intermediary GANs if necessary.}
\end{figure*}

\subsection{Evaluation procedure}

The evaluation of a GAN is not a simple task~\cite{borji2019pros,grnarova2018evaluating}. 
Fortunately, following~\cite{rodriguez2018fast}, we can evaluate the quality of the generated samples with three summary statistics commonly used in the field of cosmology.
\begin{enumerate}
    \item \emph{Mass histogram} is the average (normalized) histogram of the pixel value in the image. Note that the pixel value is proportional to the matter density.
    \item \emph{Power spectrum density} (or PSD) is the amplitude of the Fourier modes as a function of their frequency (the phase is ignored). Practically, a set of bins is defined and the modes of similar frequency are averaged together.
    \item \emph{Peak histogram} is the distribution of maxima in the density distribution, often called ``peak statistics'', or ``mass function''. Peaks capture the non-Gaussian features present in the cosmic web. This statistic is commonly used on weak lensing data \cite{kacprzak2016cosmology, martinet2017kids}. A peak is defined as a pixel greater than every pixel in its 2-pixels neighbourhood (24 pixels for 2D and 124 for 3D).
\end{enumerate}
Other statistics such as Minkowski functionals, three point correlation functions, or phase distributions, could be considered.
Nevertheless, we find that the three aforementioned statistics are currently sufficient to compare different generative models.

\paragraph{Distance between statistics}
We define a score that reflects the agreement of the 3 aforementioned cosmological measures. Problematically, the scalars forming the 3 vectors representing them have very different scales and their metrics should represent the relative error instead of the absolute one. For this reason, we first compute the logarithm (in base 10) of the computed statistics $s$.
As all statistics are positive but not strictly positive, we add a small value $\epsilon$ before computing the logarithm, i.e., $s_{\log}= \log_{10}(s+\epsilon)$.
$\epsilon$ is set to the maximum value of the statistic averaged over all real samples divided by $10^5$.

At this point, the relative difference connects to the difference of the real and fake $s_{\log}$, i.e.: ${s_{\log}^{r}-s_{\log}^f\approx \log_{10}\frac{s^r}{s^f}}$.
One could quantify the error using a norm, i.e.: ${\|\E{s_{\log}^{r}}-\E{s_{\log}^f}\|}$.
However, such a distance does not take into account second-order moments. Rather, we take inspiration from the Fréchet Inception Distance~\cite{heusel2017gans}.
We start by modeling the real and fake log statistics $s_{\log}^{r},s_{\log}^{f}$ as two multivariate Gaussian distributions. This allows us to compute the Fréchet Distance (FD) between the two distributions \cite{frechet1957distance}, which is also the Wasserstein-2. 
The FD between two Gaussian distribution with mean and covariance $(m^r,C^r)$ and $(m^f, C^f)$ is given by~\cite{dowson1982frechet}:
\begin{equation} \label{eq:frechet_distance}
    d^2\left((m^r,C^r),(m^f, C^f) \right) = \|m^r - m^f\|_2^2 + \text{Tr}\left( C^r + C^f - 2 C^r C^f\right)
\end{equation}
Note that \cite{dowson1982frechet} also proves that $\text{Tr}\left( C^r + C^f - 2 C^r C^f\right)$ is a metric for the space of covariance matrices.
We choose the FD over the Kullback-Leibler (KL) divergence for two reasons: a) the Wasserstein distance is still an appropriate distance when distributions have non-overlapping support and b) the KL is computationally more unstable since the covariance matrices need to be inverted.

\subsection{$N$-body mass map generation challenge}
\label{sec:score}
Using the FD, we define a {\it score} for each statistic as
\begin{equation}\label{eq:score}
S^* = \frac{1}{d^2((m^r,C^r),(m^f, C^f)},
\end{equation}
where $m,C$ are computed on the log statistics, i.e: {$m^r=\E{s_{\log}^{r}}$, $m^f=\E{s_{\log}^{f}}$}.
Along with this manuscript, we release our dataset and our evaluation procedure in the hope that further contributions will improve our solution. All information can be found at \url{https://github.com/nperraud/3DcosmoGAN}.
We hope that this dataset and evaluation procedure can be a tool for the evaluation of GANs in general.
Practically, these cosmological statistics are very sensitive.
We observed two important properties of this problem.
First, a small variation in the generated images still has an impact on the statistics.
The statistics can be highly affected by high density regions of the $N$-body data, and these regions are also the most rare in the training set.
Second, while mode collapse may not directly affect the mean of the statistics, it can affect their second order moment significantly.
We observed that obtaining a good statistical agreement (and hence a high score), is much more difficult than obtaining generated images that are indistinguishable for the human eye, especially for the 2-dimensional case.
We found that the problems of high data volume, large dynamic range of the data, and strict requirement on good agreement in statistical properties of real and generated samples, pose significant challenge for generative models.

\paragraph{Interpretation of the $S^*$ scores}
We report the $S^*$ scores for the 3D case in Section~\ref{sec:experiments} and the web page hosting the source code also includes baseline scores for the 2D case.
We would like to emphasize that these scores are mostly suitable to compare two generative models applied to the \emph{same} distribution. It is a priori unclear whether these scores provide a meaningful comparison when considering two generative models trained on different distributions. We therefore refrain from relying on such scores to compare the generative models trained on 2D and 3D data since the latter gives access to correlations along the third dimension that are absent from the 2D data.
Finally, while in theory, the $S^*$ score is unbounded and can be arbitrarily large as $d^2\rightarrow 0$, its empirical evaluation is, in practice, limited by the estimation error of $d^2$ that depends on the number of samples used to estimate the mean vectors and the covariance matrices $(m^r,C^r)$ as well as the moments of the estimated statistics.
In the case where the training set contains a large number of samples, one would expect the training score to be indicative of the score of the true distribution. One can see that the estimation error of the first term of \eqref{eq:frechet_distance} depends mostly on the variance of the statistic, while for the second term, it depends on the moments of order $3$ and $4$. Hence, the estimated score will reflect the variance of the estimated statistics given the number of samples used. A high score therefore means less variance within the corresponding statistic.

\begin{figure*}[ht!]
    \centering
    \final{
}{
\includegraphics[width=0.3\textwidth]{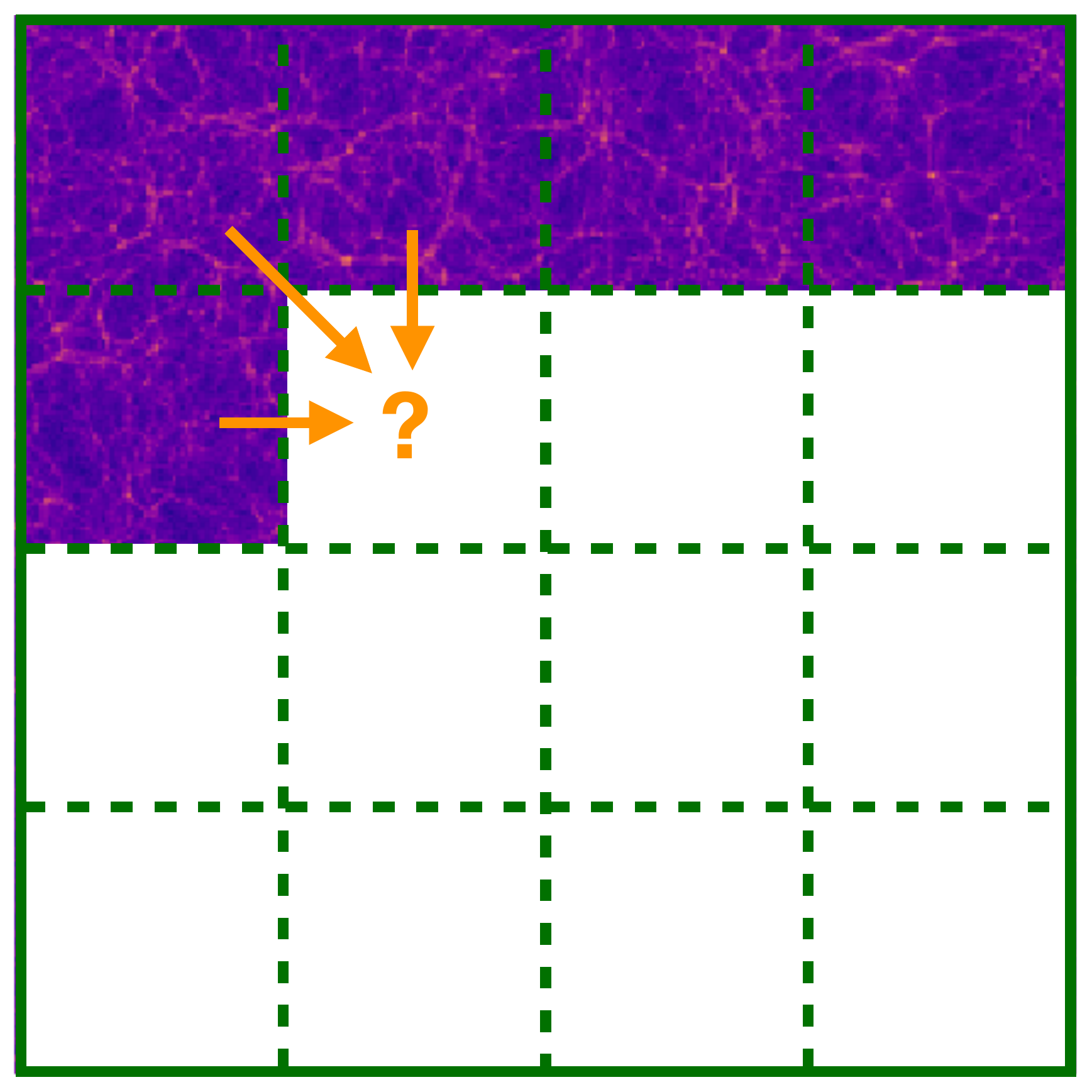}\hspace{1cm}
\includegraphics[width=0.6\textwidth]{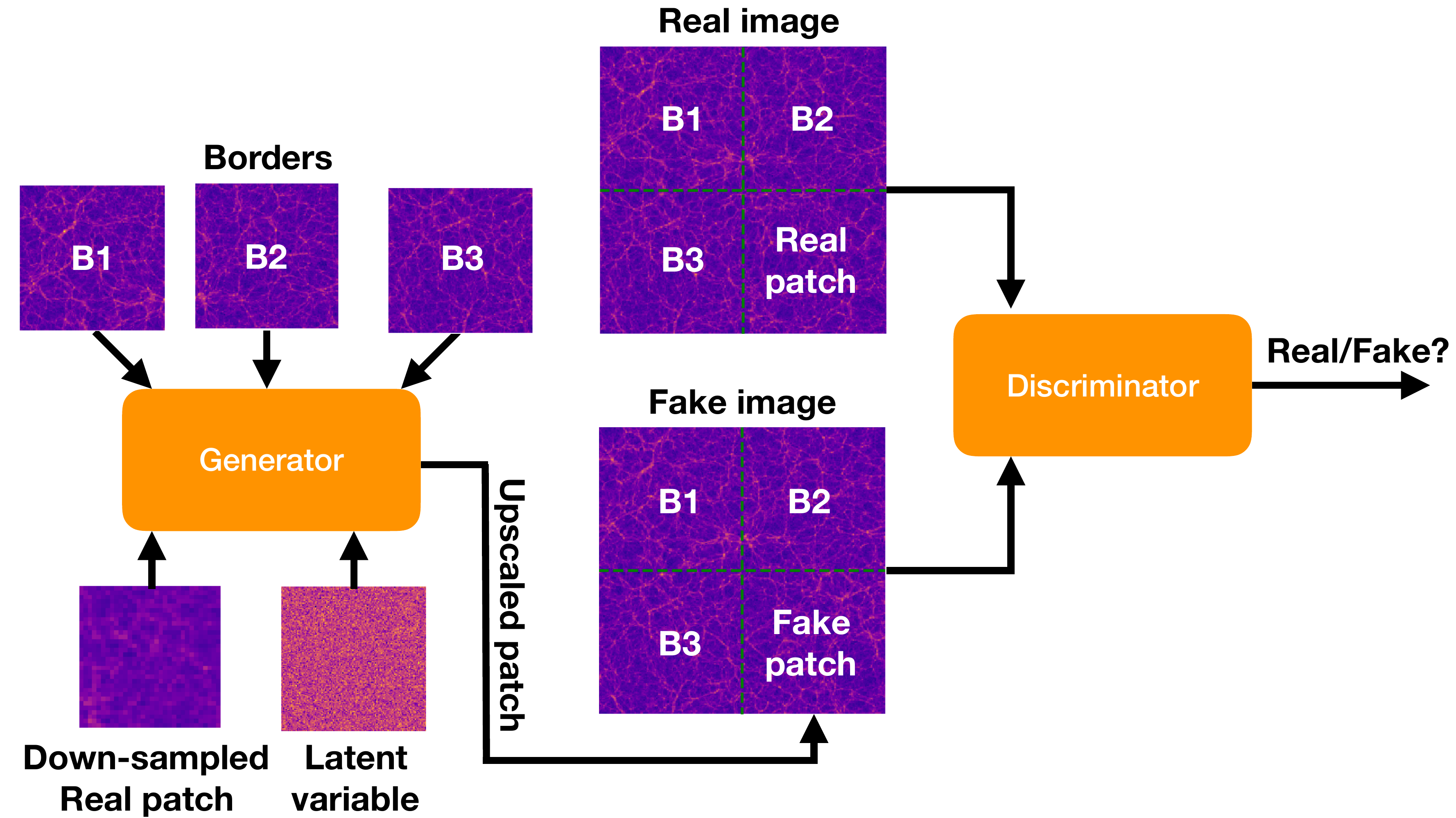}}
\caption{\label{fig:Multi-Scale-Model}Left: Sequential Generation. The image is generated patch-by-patch. Right: Upscaling GAN conditioned on neighborhood information. Given border patches B1, B2 and B3, and the down-scaled version of the $4^{th}$ patch, the generator generates the $4^{th}$ fake patch. The discriminator also receives the border patches, as well as either an original or fake patch from the generator. The principle is similar in 3D, with the difference that $7$ border cubes are used instead of $3$ border patches.}
\end{figure*}
\begin{figure}[ht!]
    \centering
    \final{
    }{
\includegraphics[width=0.45\textwidth]{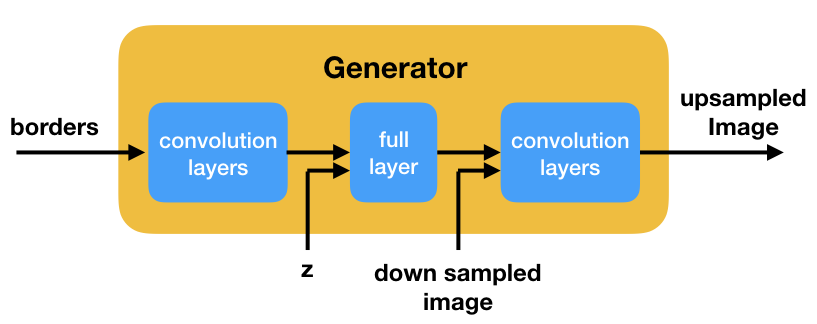}
\hspace{1cm}
}
\caption{\label{fig:generator-details} Details of the upsampling generator. The borders/cubes are encoded using several convolutional layers.}
\end{figure}

\section{Sequential generative approach}
\label{sec:model}
We propose a novel approach to efficiently learn a Generative Adversarial Network model (see Section~\ref{sec:gan}) for \mbox{2D} and \mbox{3D} images of arbitrary size. Our method relies on two building blocks: 1) a multi-scale model that improves the quality of the generated samples, both visually and quantitatively, by learning unique features at different scales (see Section~\ref{sec:multiscale}), and 2) a training strategy that enables learning images of arbitrary size, that we call ``conditioning on neighborhood'' (see Section~\ref{sec:conditioning}).

\subsection{Generative Adversarial Networks (GANs)}
\label{sec:gan}
Generative Adversarial Networks (GAN) rely on two competing neural networks that are trained simultaneously: the generator $\mathrm{G}$, which produces new samples, and the discriminator $\mathrm{D}$, which attempts to distinguish them from the real ones. 
During training, it is the generator's objective to fool the discriminator, while the discriminator resists by learning to accurately discriminate real and fake data. 
Eventually, if the optimization process is carried out successfully, the generator should improve to the point that its generated samples become indistinguishable from the real one.
In practice, this optimization process is challenging and numerous variants of the original GAN approach have been proposed, many of them aiming to improve stability including e.g.~\cite{roth2017stabilizing, gulrajani2017improved, arjovsky2017wasserstein}. In our work, we rely on the improved Wasserstein GAN (WGAN) approach introduced in~\cite{gulrajani2017improved}. The latter optimizes the Wasserstein distance instead of the Jensen-Shannon divergence and penalizes the norm of gradient of the critic instead of using a hard clipping as in the original WGAN~\cite{arjovsky2017wasserstein}. The resulting objective function is
 \begin{equation*}
    \min_G \max_{D \in \mathcal{D}} \mathop{\mathbb{E}}_{\pmb{x} \sim \mathbb{P}_r}[D(\pmb{x})] - \mathop{\mathbb{E}}_{\widetilde{\pmb{x}} \sim \mathbb{P}_g}[D(\widetilde{\pmb{x}})] + \lambda \mathop{\mathbb{E}}_{\hat{\pmb{x}} \sim \mathbb{P}_{\hat{\pmb{x}}}}[( || \nabla_{\hat{\pmb{x}}} D(\hat{\pmb{x}} )||_{2} - 1)^2],
\end{equation*}
where $\mathbb{P}_r$ is the data distribution and $\mathbb{P}_g$ is the generator distribution implicitly defined by $\widetilde{\pmb{x}} = G(\mathrm{z}), \mathrm{z} \sim p(\mathrm{z})$. The latent variable $\mathrm{z}$ is sampled from a prior distribution $p$, typically a uniform or a Gaussian distribution.
Eventually, $\mathbb{P}_{\hat{\pmb{x}}}$ is defined implicitly by sampling uniformly along straight lines between pair of points sampled from the true data distribution $\mathbb{P}_r$ and the generator distribution $\mathbb{P}_g$. The weight $\lambda$ is the penalty coefficient.

\subsection{Multi-scale Model}
\label{sec:multiscale}
Our multi-scale approach is inspired by the Laplacian pyramid GAN~\cite{LAP_GAN}.
We refer to three image types of different sizes, namely $I_3 = 32 \times 32 \times 32, I_2 = 64 \times 64 \times 64, I_1 = 256 \times 256 \times 256$~\footnote{We will use the abbreviations $32^3, 64^3$ and $256^3$ for conciseness.} pixels, where $I_2$ is a down-scaled version of $I_1$ and $I_3$ is a down-scaled version of $I_2$.
The multi-scale approach is shown in figure \ref{fig:Multi-Scale_Overview} and uses three different GANs, $M_1$, $M_2$ and $M_3$, all trained independently from each other, and can therefore be trained in parallel.
We train GAN $M_3$ to learn the data distribution of images $I_3$, while the GANs $M_2$ and $M_1$ are conditioned on the images produced by $M_3$ and $M_2$, respectively.
In our implementation, we take $M_3$ to be a normal Deep Convolution GAN (\mbox{DCGAN}) that learns to produce down-scaled samples of size $I_3$.
We design GANs $M_2$ and $M_1$ to have the following properties:
1) they produce outputs using a sequential patch-by-patch approach and
2) the output of each patch is conditioned on the neighboring patches.
This procedure allows handling of high data volume, while preserving the long-range features in the data.
Moreover, different GANs learn salient features at different scales, which contribute to an overall improved quality of the samples produced the final GAN $M_1$.
Further details regarding the implementation details are provided in Appendix~\ref{sec:architecture_details}.

\subsection{Conditioning on Neighborhoods}
\label{sec:conditioning}
The limited amount of memory available to train a GAN generator makes it impractical to directly produce large image samples. Using current modern GPUs with 16GB of RAM and a state-of-the-art network architecture, the maximum sample size we were allowed to use was $32^3$, which is far from our target taken to be $256^3$. In order to circumvent this limitation, we propose a new approach that produces the full image (of size $256^3$) patch-by-patch, each patch being of smaller size ($32^3$ in our case). This approach is reminiscent of the Pixel-CNN \cite{Pixel_CNN}, where \mbox{2D} images are generated pixel-by-pixel, rather than the entire picture being generated at once. Instead of assuming that the distribution of each patch is independent of the surrounding context, we model it as a function of the neighboring patches. The generation process is done using a raster-scan order, which implies that a patch depends on the neighboring patches produced before the current patch. The process illustrated in Figure~\ref{fig:Multi-Scale-Model} is for the \mbox{2D} case with three neighboring patches; the generalization to three dimensions is straightforward as it simply requires seven neighboring \mbox{3D} patches.

\begin{figure*}[th!]
    \centering
    \final{
    }
    {
    \includegraphics[width=0.4\textwidth]{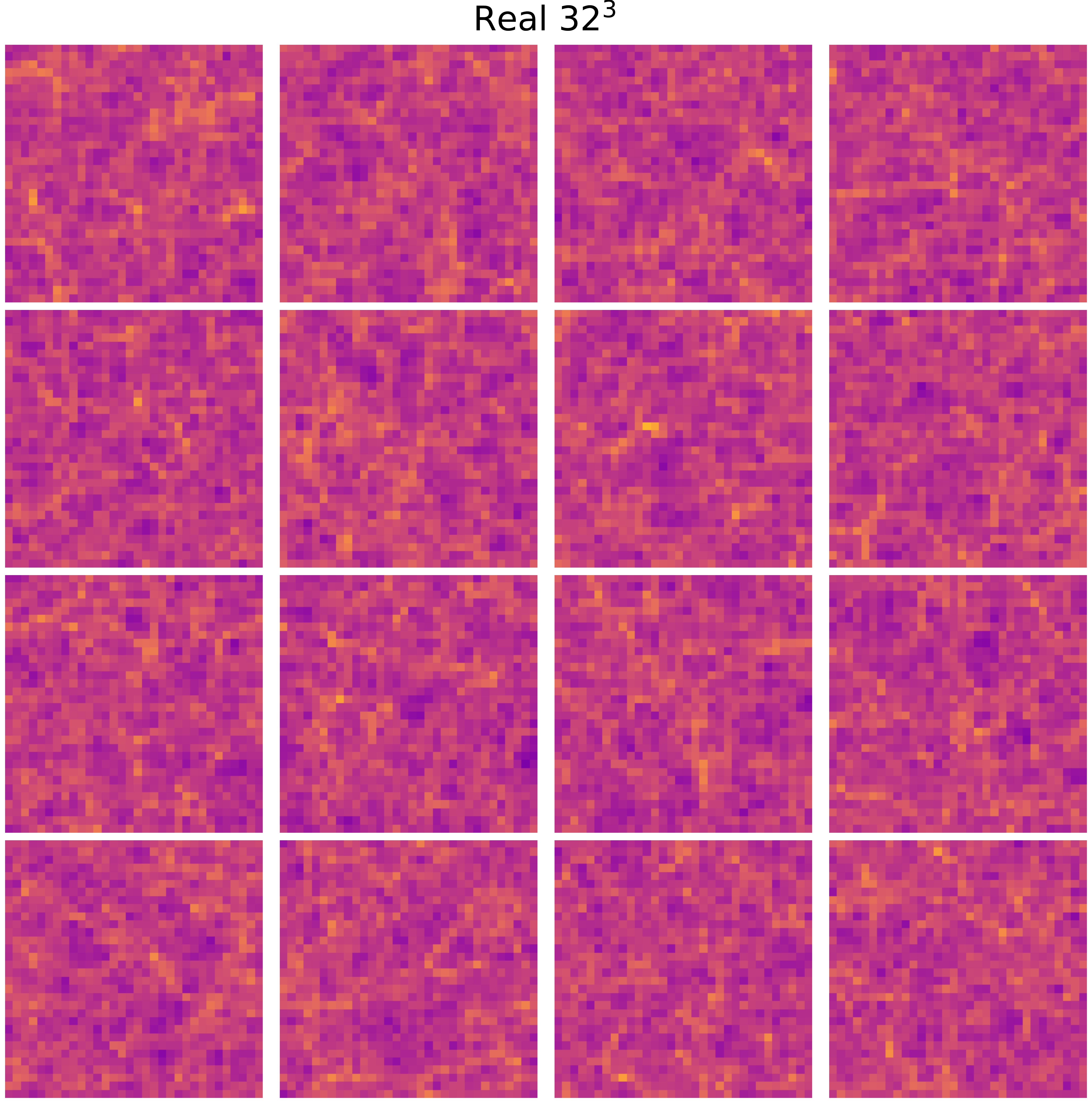}
    \hspace{0.5cm}
    \includegraphics[width=0.4\textwidth]{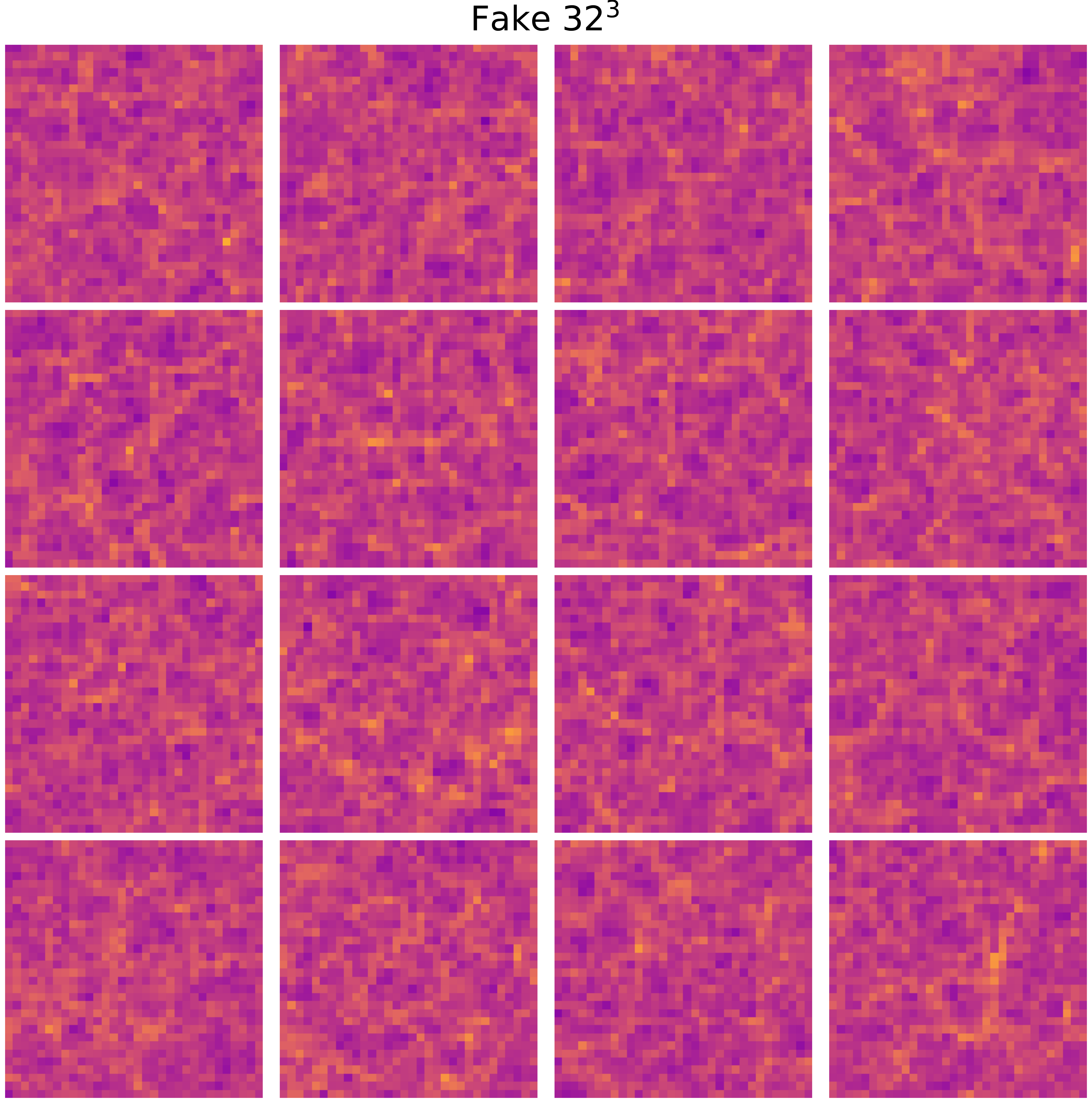}
    }
    \caption{ \label{fig:32_slices} $8$-down-scaled sample generation ($32^3$ cubes). Middle slice from $16$ real and $16$ fake WGAN $M_3$ samples.
    Video: \url{https://youtu.be/uLwrF73wX2w}
    }
    
    \centering
    \final{
    }
    {
    \includegraphics[width=0.27\textwidth]{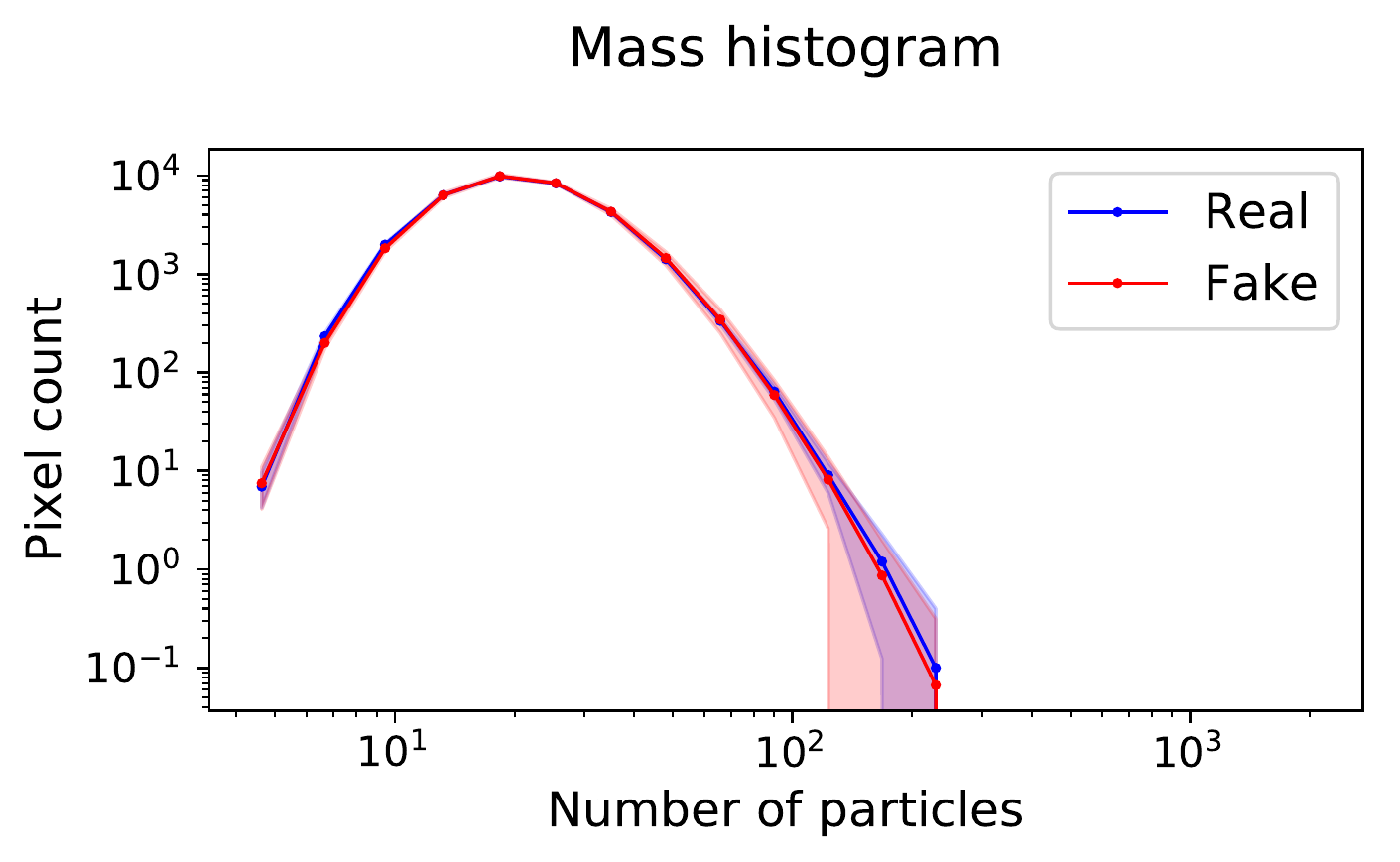}
    \includegraphics[width=0.27\textwidth]{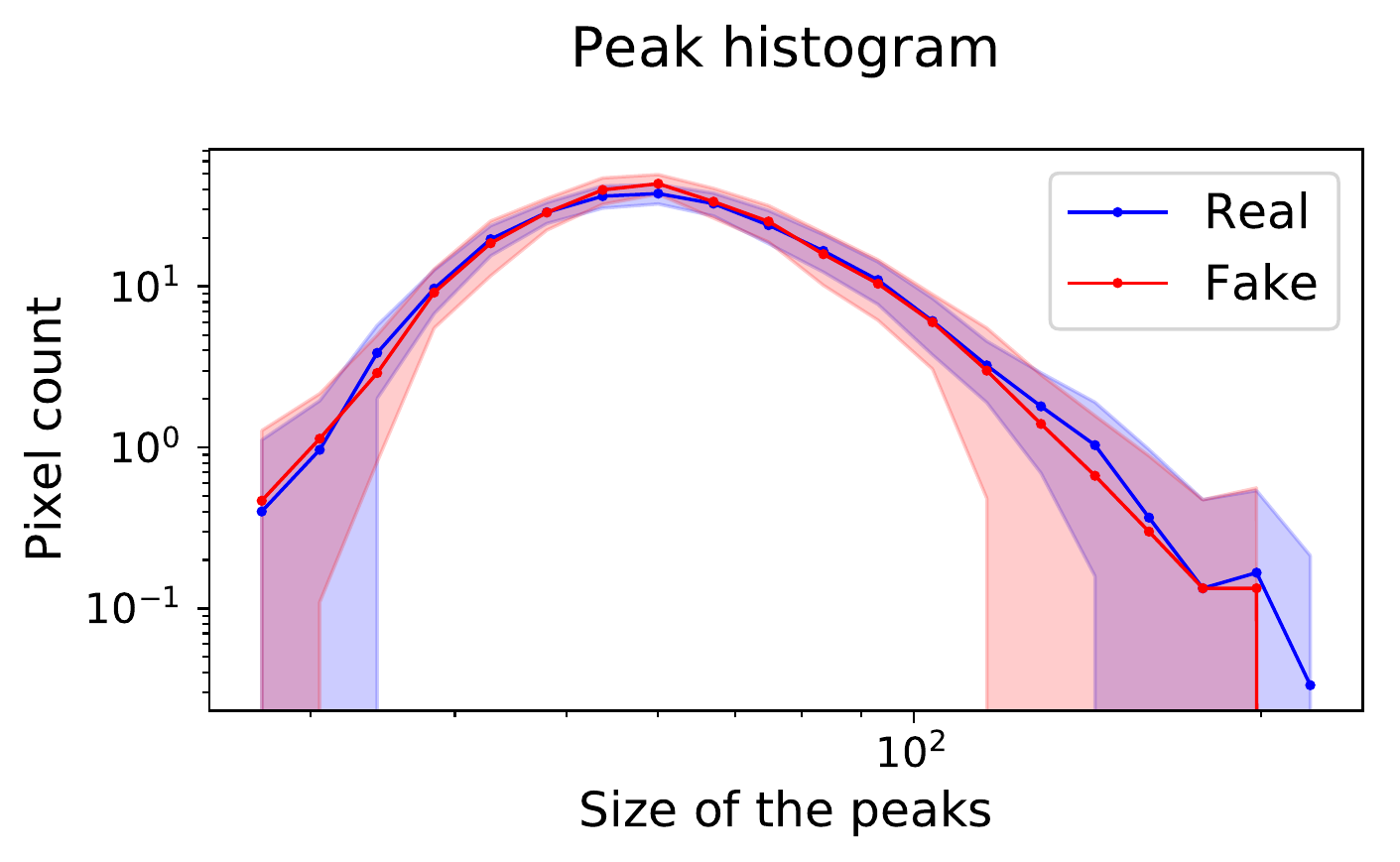}
    \includegraphics[width=0.27\textwidth]{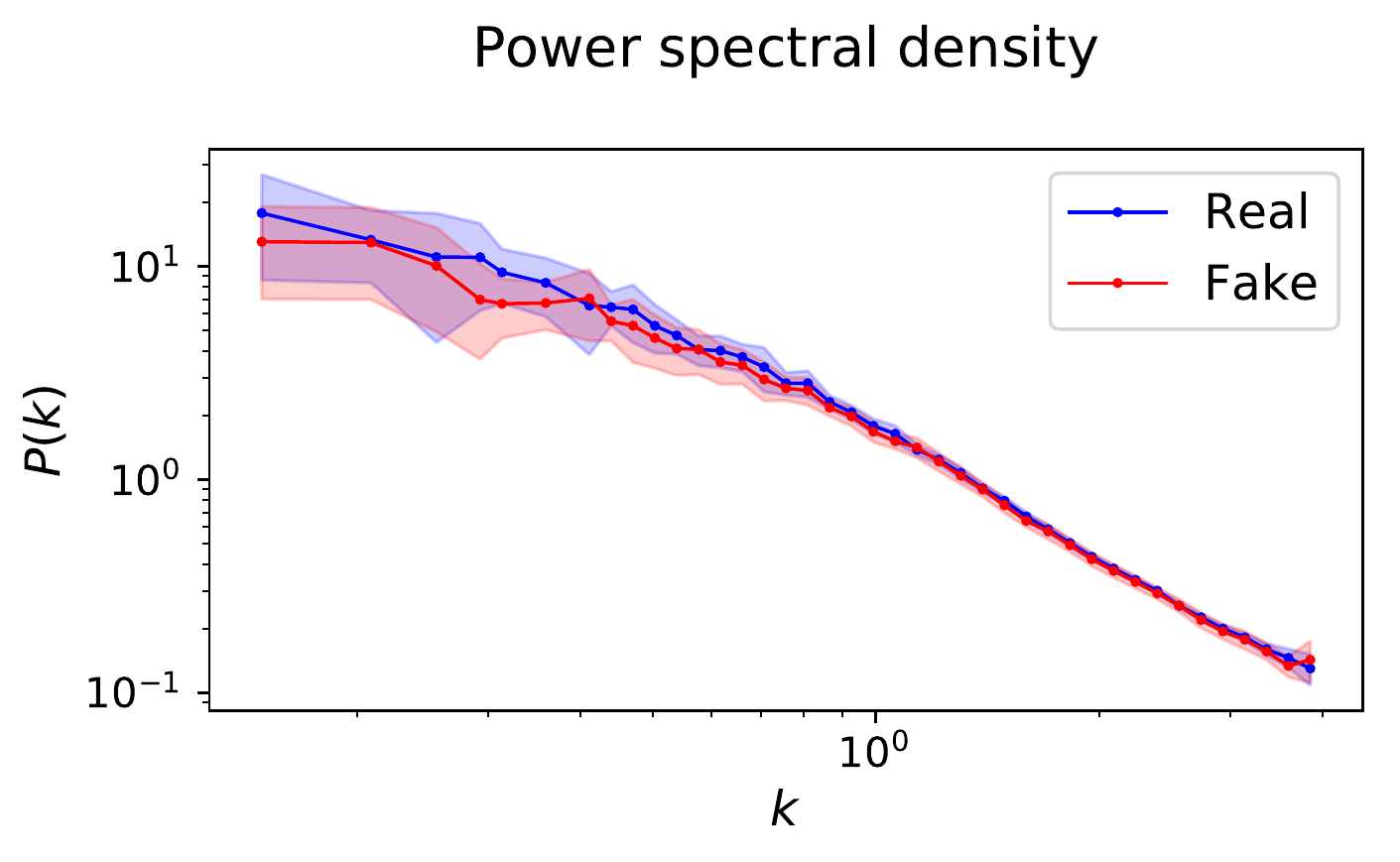}
    }
    \caption{\label{fig:32_stat} Statistics of the $8$-downscaled $32^3$ cubes. The fake samples are generated from WGAN $M_3$. The power spectrum density is shown in units of h Mpc$^{-1}$, where h = H$_0$/100 corresponds to the Hubble parameter.}
    
  \vspace{0.25cm}
  \hrulefill
  \vspace{0.25cm}
  
    \centering
    \final{
    }
    {
    \includegraphics[width=0.27\textwidth]{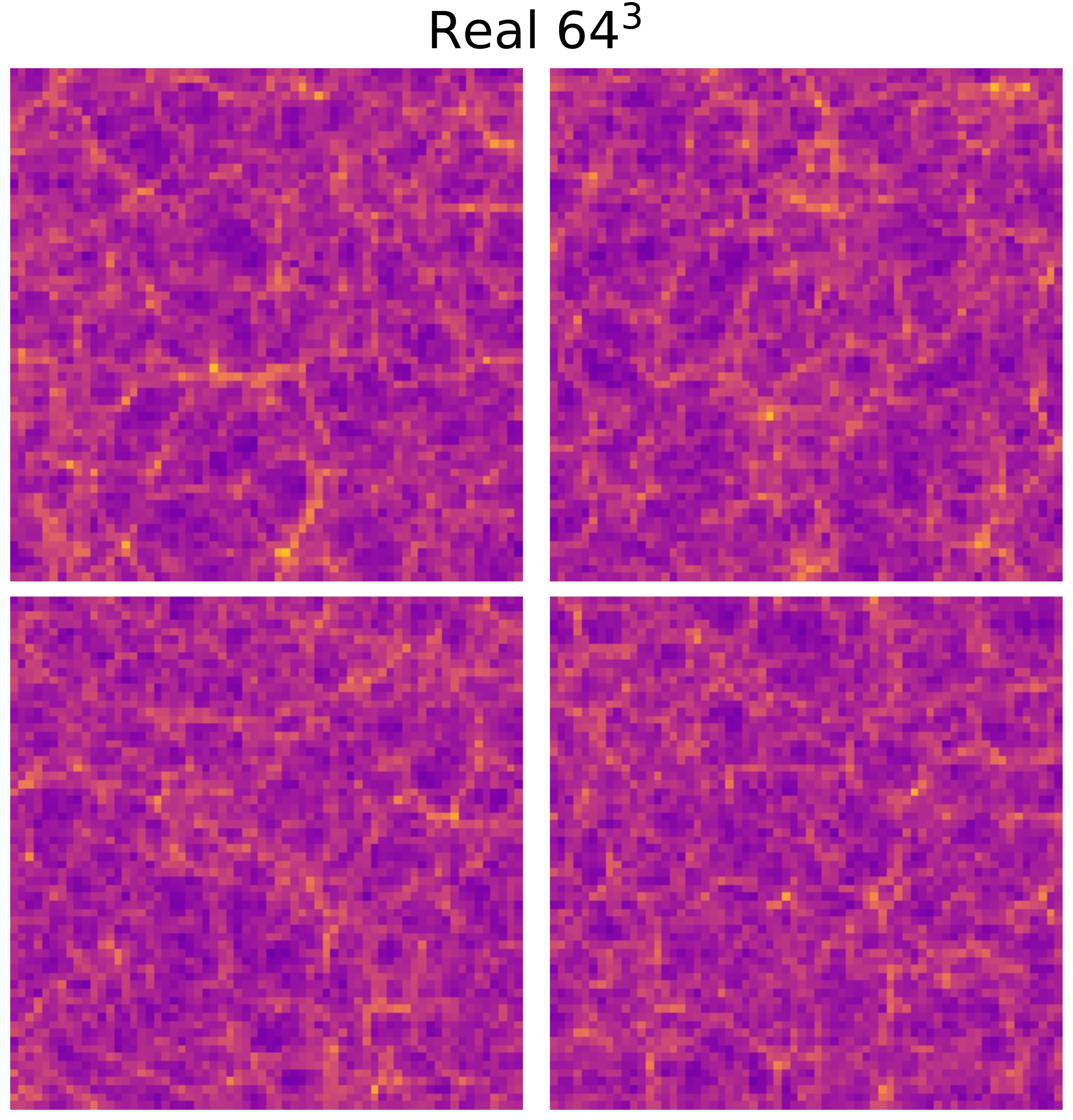}
    \hspace{0.25cm}
    \includegraphics[width=0.27\textwidth]{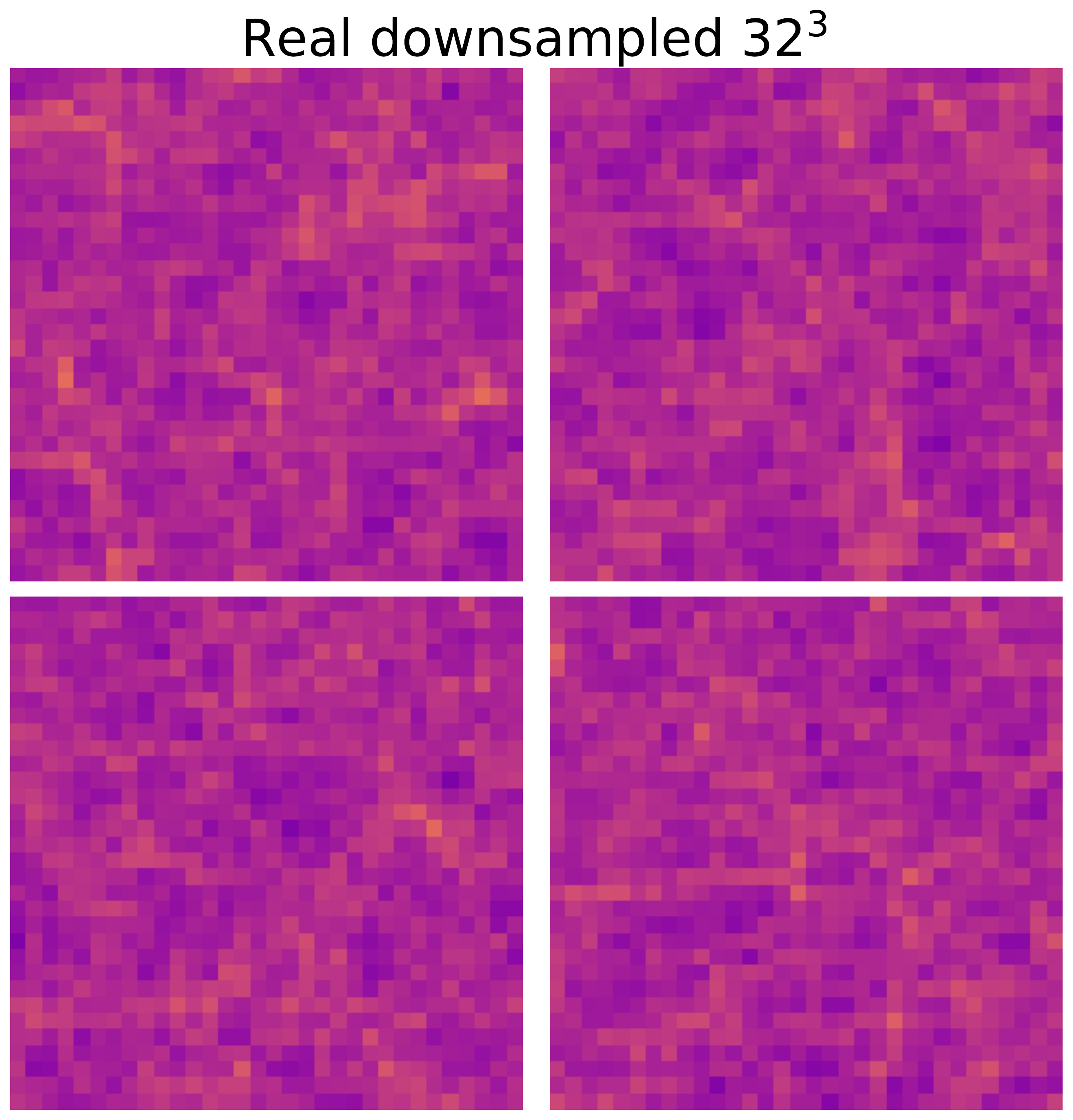}
    \hspace{0.25cm}
    \includegraphics[width=0.27\textwidth]{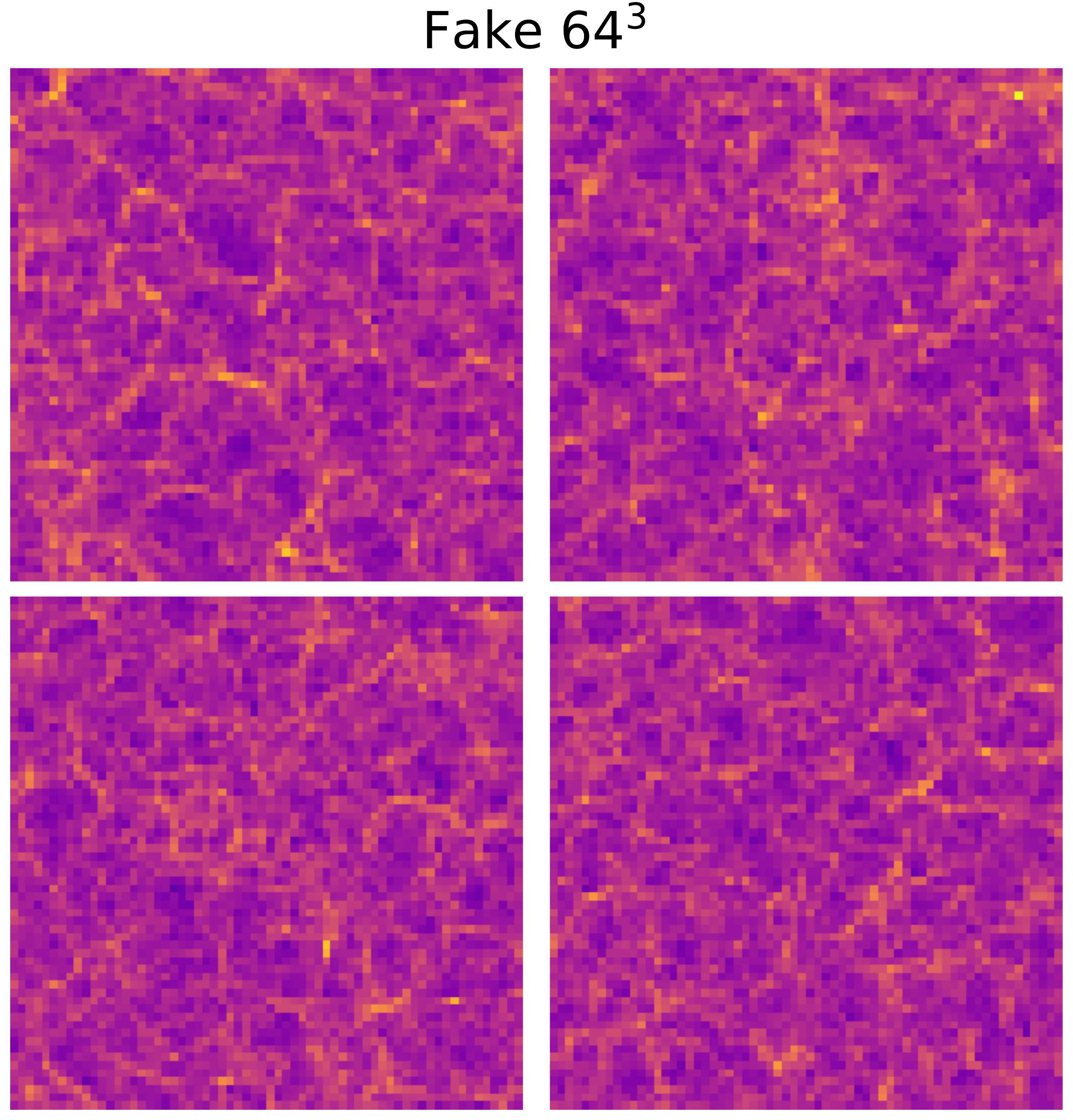}}
    \caption{\label{fig:64_slices} Up-scaling a $32^3$ cube to $64^3$. Left and right: middle slices from $4$ real and fake $64^3$ samples. The fake is generated by conditioning the WGAN $M_2$ on the real down-scaled $32^3$ cube (center).
    Video: \url{https://youtu.be/IIPoK8slShU}
    }
    \centering
    \final{
    }{
    \includegraphics[width=0.27\textwidth]{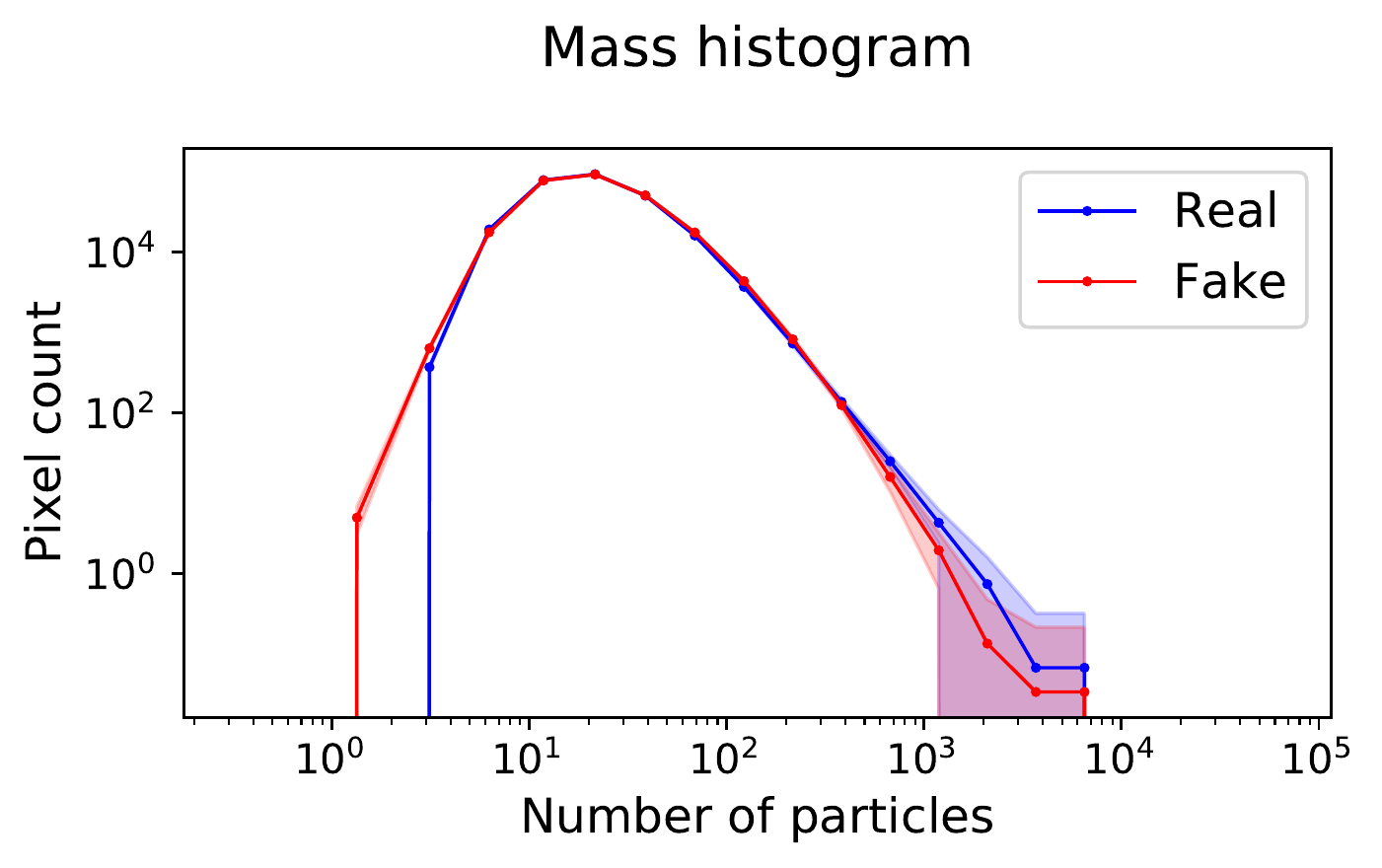}
    \includegraphics[width=0.27\textwidth]{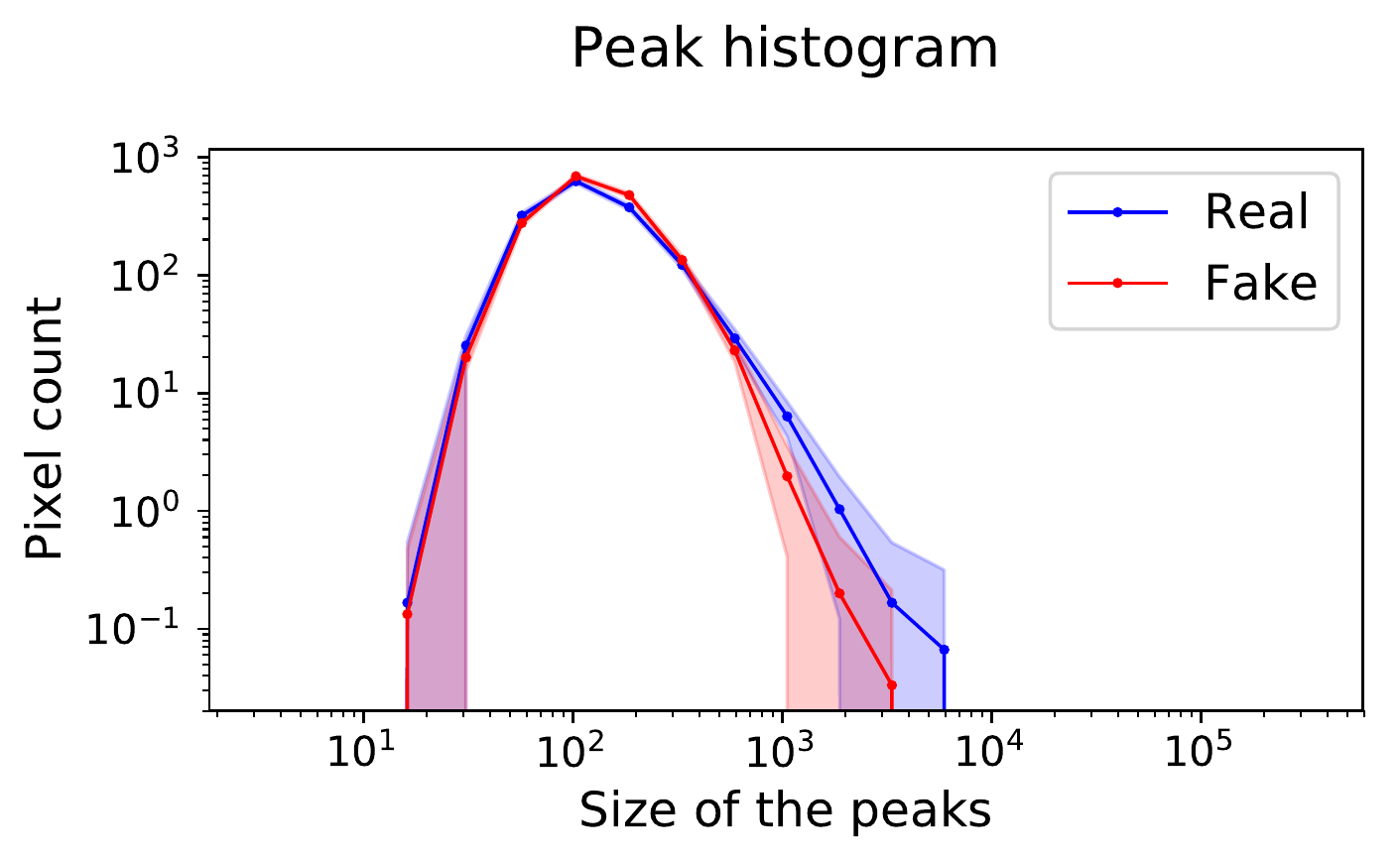}
    \includegraphics[width=0.27\textwidth]{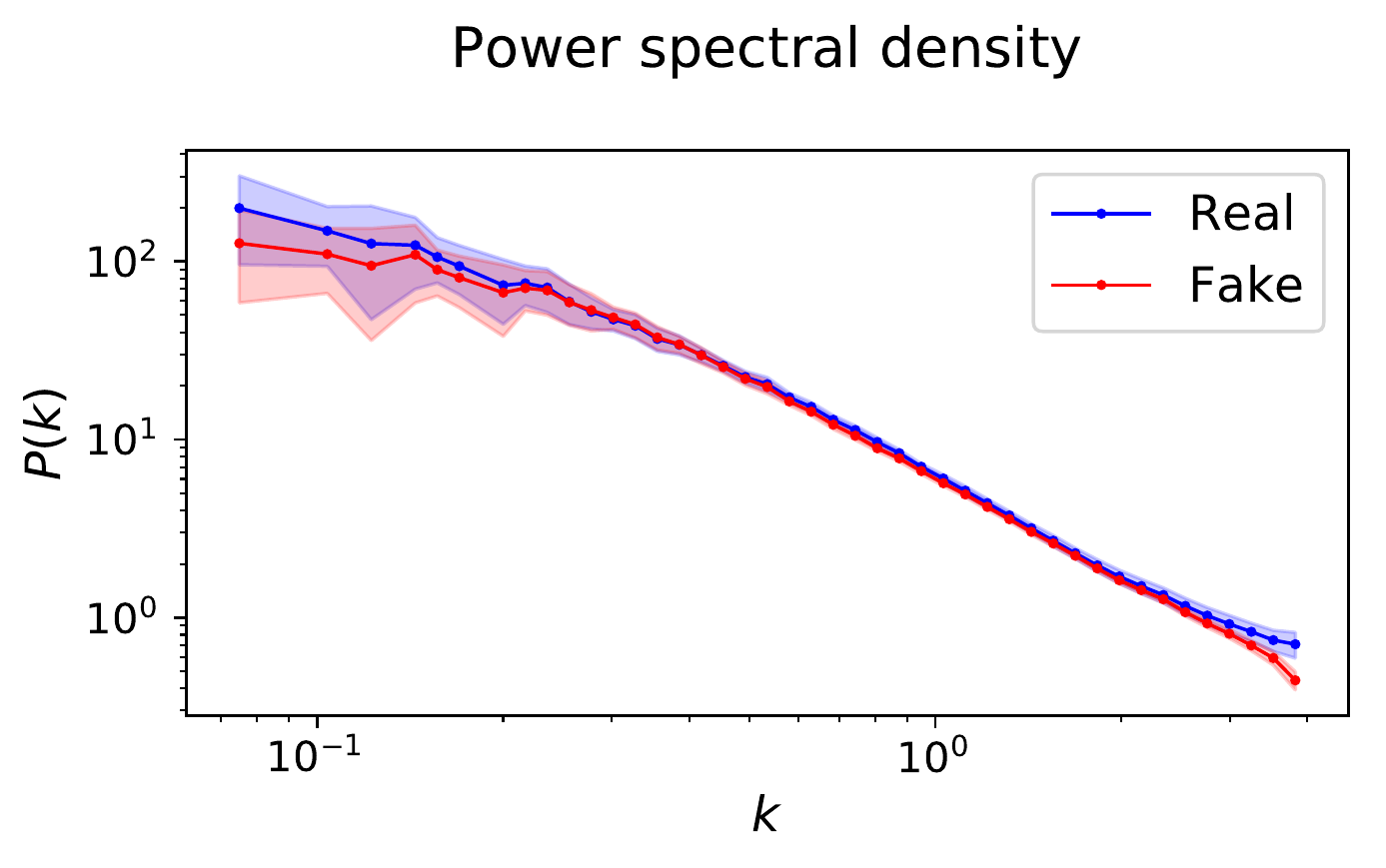}}
    \caption{\label{fig:64_stat} Statistics of the samples produced by $M_2$. The fake samples are generated by conditioning WGAN $M_2$ on the real down-scaled $32^3$ samples. The power spectrum density is shown in units of h Mpc$^{-1}$, where h = H$_0$/100 corresponds to the Hubble parameter.}
\end{figure*}

In the generator, the neighborhood information, i.e. the borders ($3$ patches for 2D, $7$ cubes for 3D), is first encoded using several convolutional layers. Then it is concatenated with the latent variable, is inputed to a fully connected layer before being reshaped into a 3D image. At this point, the down-sampled version of the image is concatenated. After a few extra convolutional layers, the generator produces the lowest rightmost patch with the aim of making it indistinguishable to the patch from the real data. The generator architecture is detailed in Figure~\ref{fig:generator-details}. In the case where there is no neighborhood information available, such as at the boundary of a cube, we pad with zeros.
The discriminator is given images containing four patches where the lower right patch is either the real data or the fake data generated by the generator.
The generator only produces patches of size $32^3$, irrespective of the size of the original image. This way this method can scale to any image size, which is a great advantage. The discriminator only has access to a limited part of the image and ignores the long-range dependencies. This issue, however, is handled by the multi-scale approach described in the previous section.
We actually tried a model only conditioned on the neighborhoods as detailed in Appendix~\ref{sec:single-scale-model}. It ended up performing significantly worse than using the multi-scale approach.

\begin{figure*}[th!]
    \centering
    \final{
    }
    {
    \includegraphics[width=0.3\textwidth]{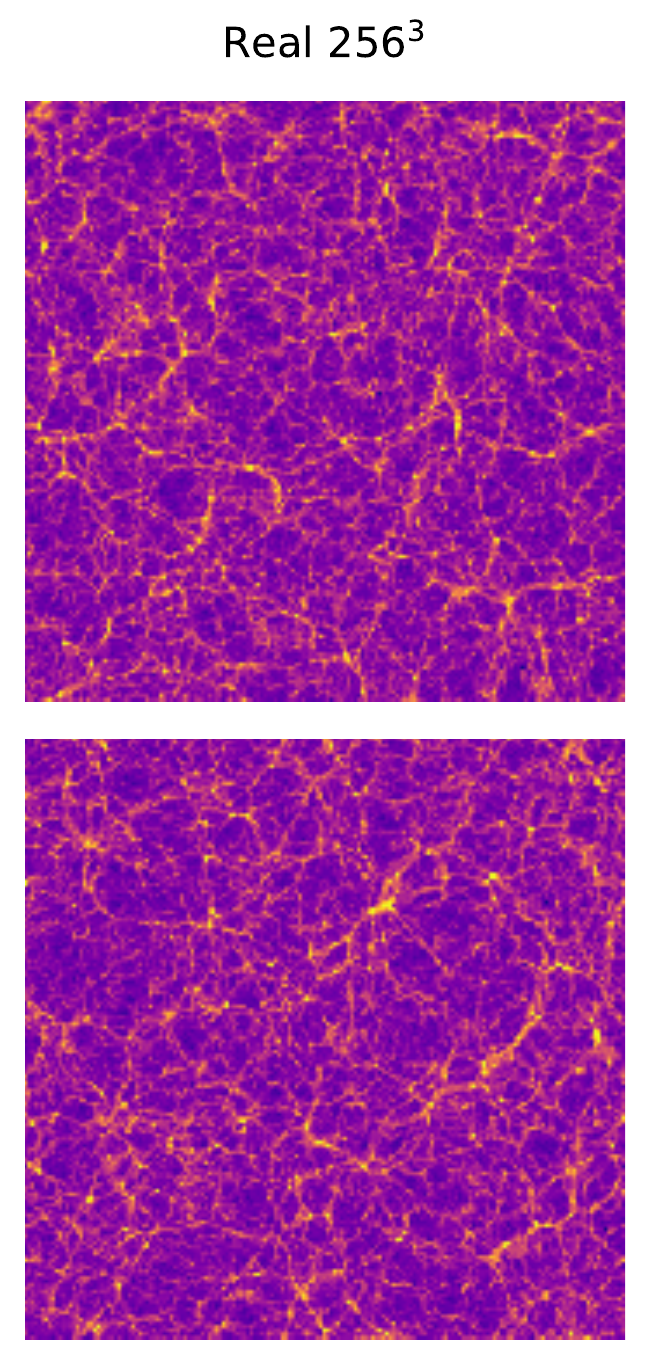}
    \hspace{0.1cm}
    \includegraphics[width=0.3\textwidth]{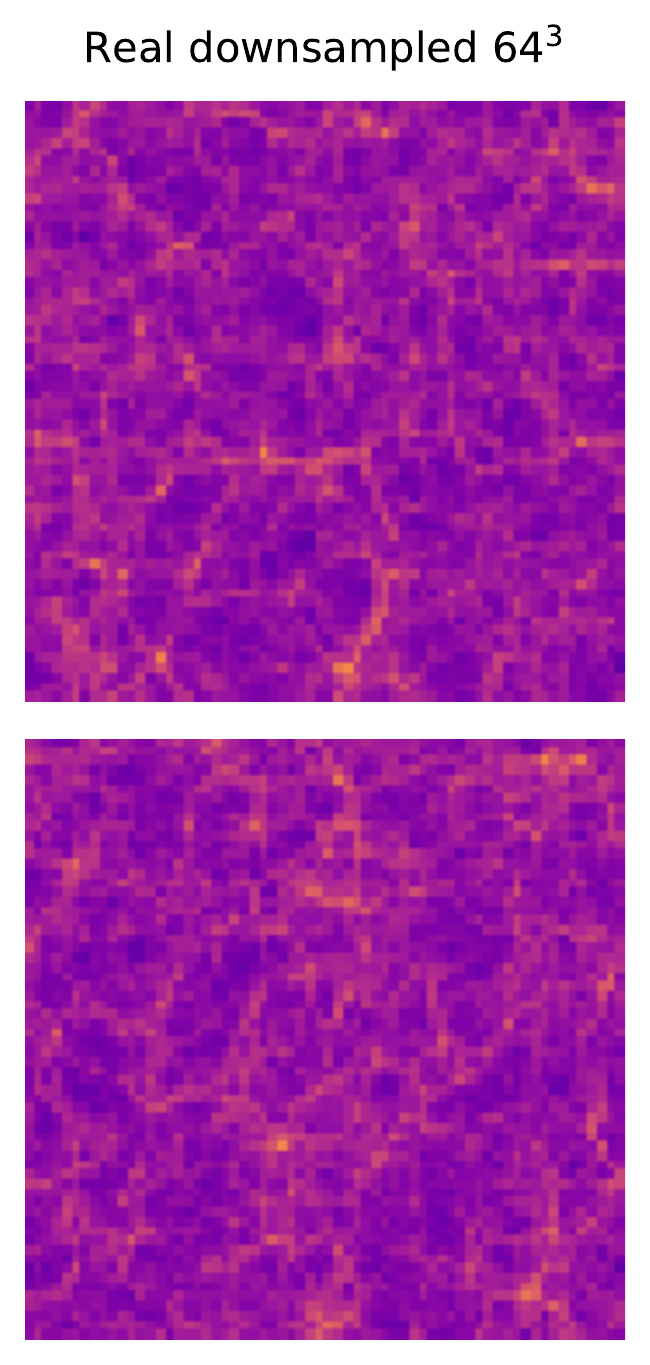}
    \hspace{0.1cm}
    \includegraphics[width=0.3\textwidth]{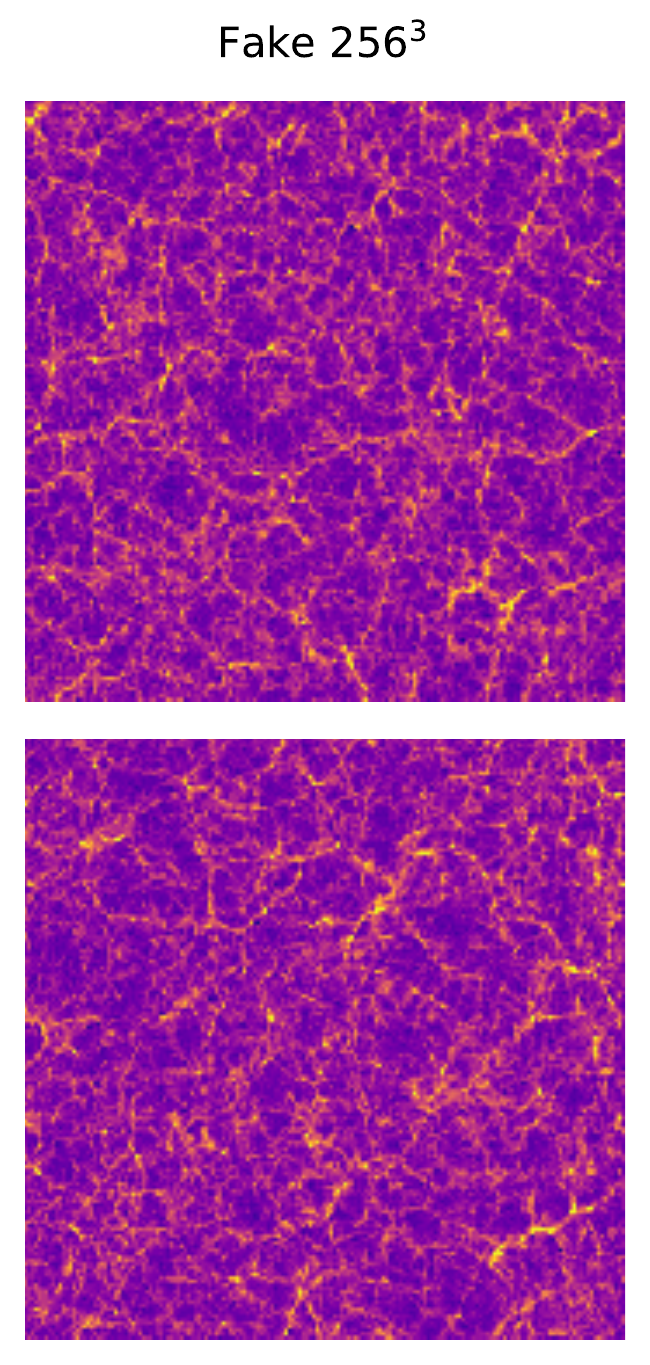}
    }
    \caption{\small{Upsampling a $64^3$ cube to $256^3$. Left and right: middle slices from $2$ real and fake $256^3$ samples. The WGAN $M_1$ that generates the fake sample is conditioned on the real image down-scaled to $64^3$ (center).}
    Video: \url{https://youtu.be/guUYP8ZOoVU}
    }
    \label{fig:256_slices}
    \centering
    \final{
    }
    {
    \includegraphics[width=0.3\textwidth]{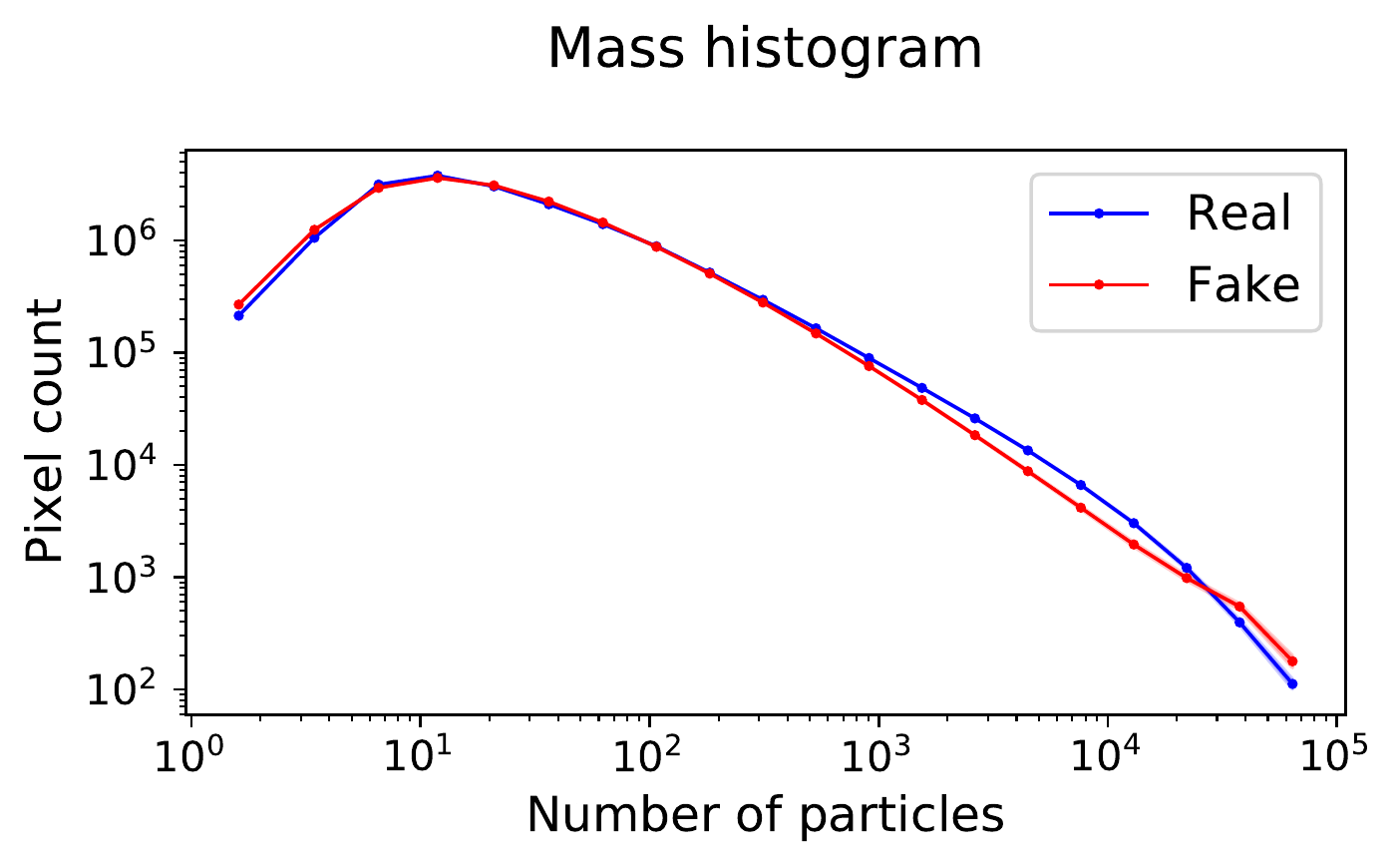}
    \includegraphics[width=0.3\textwidth]{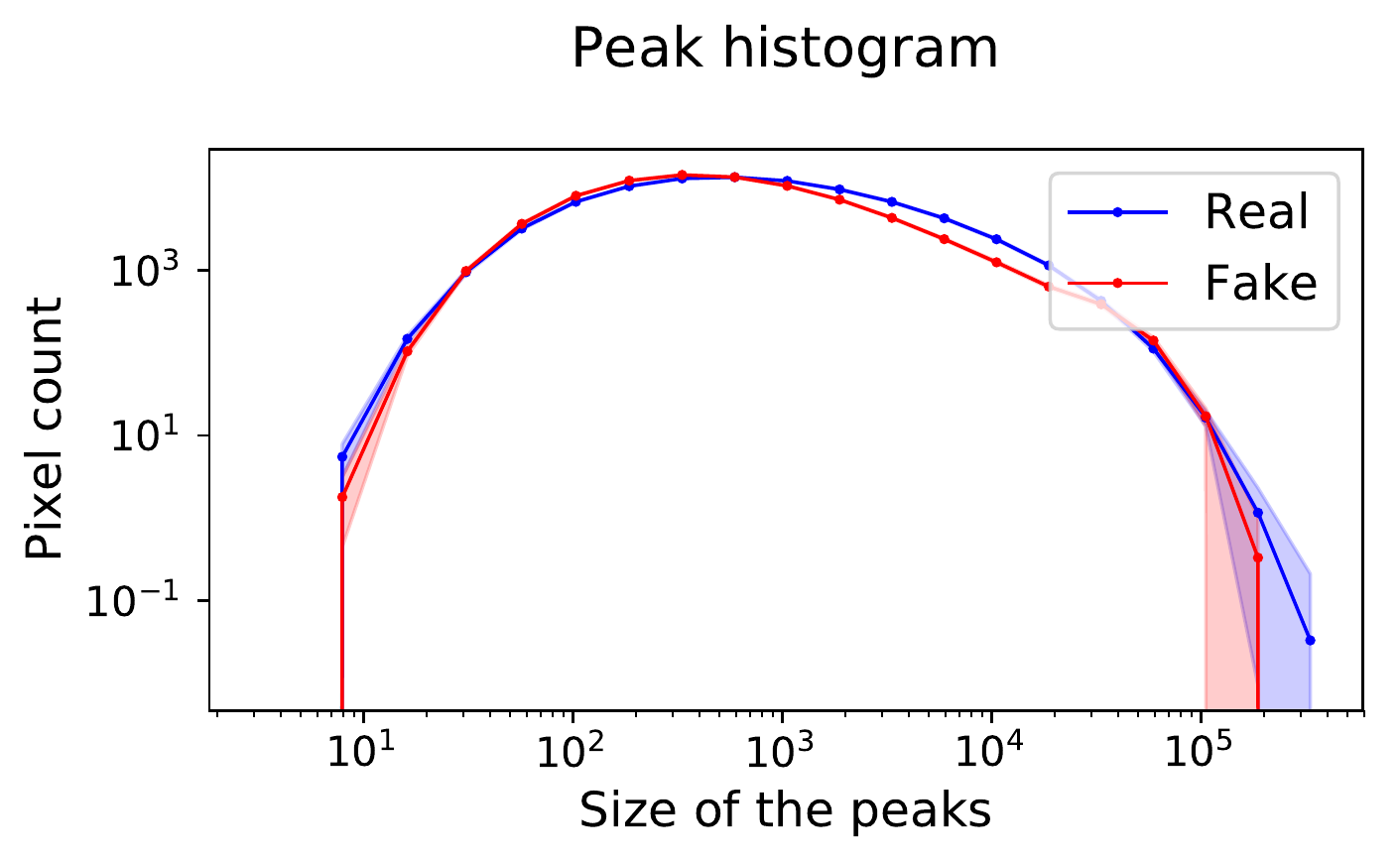}
    \includegraphics[width=0.3\textwidth]{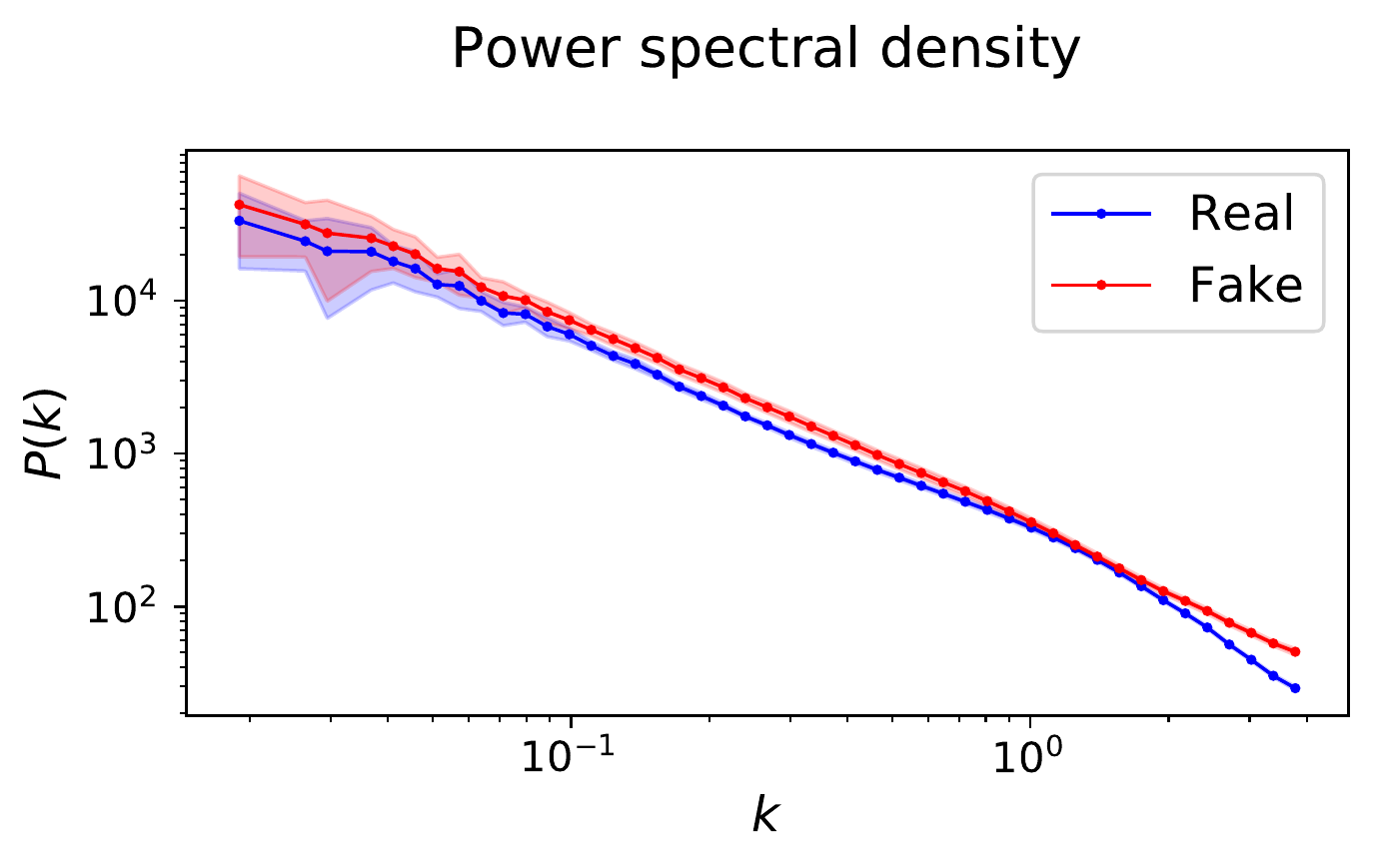}
    }
    \caption{\small{Statistics for the WGAN $M_1$ producing fake $256^3$ cubes. The fake samples are generated by conditioning WGAN $M_1$ on the real cube down-scaled to $64^3$. The power spectrum density is shown in units of h Mpc$^{-1}$, where h = H$_0$/100 corresponds to the Hubble parameter.}}
    \label{fig:256_stat}
\end{figure*}

\section{Experimental results}
\label{sec:experiments}

Our approach relies on a recursive approach that progressively builds a 3D cube, starting from a low-resolution and upsampling it at every step. We detail and analyze each step separately in order to understand the impact of each of these steps on the final performance of the full model. We also compare the results of our multi-scale approach to a simpler uni-scale model. Additional details regarding the network architectures and  various implementation choices are available in Appendix~\ref{sec:architecture_details}.

\subsection{Scale by scale analysis of the pipeline}

In the following, we describe our model that relies on three different GANs, namely $M_1, M_2$ and $M_3$, to generate samples at distinct resolutions. We detail each step of the generation process below.

\paragraph{Step 1: Low-scale generation (Latent Code to $I_3$)}
The low scale generation of a sample of size $32^3$ is performed by the WGAN $M_3$. The architecture of both the generator and the discriminator is composed a $5$ \mbox{3D} convolutional layers with kernels of size ${4 \times 4 \times 4}$. We use leaky ReLu non-linearity. Further details can be found in Table~\ref{tab:archi32} of the appendix.

Figure~\ref{fig:32_slices} shows the middle slice from $16$ \mbox{3D} samples $I_3$ drawn from the generator of $M_3$ compared to  real samples.
In Figure~\ref{fig:32_stat}, we additionally plot our evaluation statistics for the 30 samples corresponding to the total number of $N$-body simulations used to build the dataset.
The generated samples drawn from $M_3$ are generally similar to the true data, both from a visual and from a statistical point of view, although one can observe slight disagreements at higher frequencies.
Note that the low number of samples and the size of each of them does not allow us to compute very accurate statistics, hence limiting our evaluation.

\paragraph{Step 2: Up-scale ($I_3$ to $I_2$)}
Upscaling the sample size from $32^3$ to $64^3$ is performed by the WGAN $M_2$. The architecture is similar to $M_3$ (see Table~\ref{tab:archi64} in the appendix), 
except that the border patches are first encoded using three 3D convolutional layers and then concatenated with the latent variables before being fed to the generator.

In order to visualize the quality of up-scaling achieved by this first up-sampling step independently from the rest of the pipeline, we first down-scale the real $256^3$ samples to $32^3$, and then provide them as input $I_3$ to the WGAN $M_2$. We then observe the result of the up-scaling to $64^3$.
Figure \ref{fig:64_slices} shows slices from some generated $I_2$ samples, as well as the real down-scaled $I_3$ image and the real $I_2$ image.
We observe a clear resemblance between the up-scaled fake samples and the real samples.
The statistics for this step of the generation process are shown in Figure~\ref{fig:64_stat}. 
We observe more significant discrepancies for features that rarely occur in the training set such as for large peaks. This is however not surprising as learning from few examples is in general a difficult problem.

\paragraph{Step 3: Up-scale ($I_2$ to $I_1$)}
The final upscaling from $64^3$ to $256^3$ is performed by the WGAN $M_1$.
The architecture of both the generator and the discriminator of $M_1$ is composed of eight \mbox{3D} convolutional layers with inception modules~\cite{InceptionNet}. The inception modules have filters of three different sizes: $1 \times 1 \times 1$, $2 \times 2 \times 2$ and $4 \times 4 \times 4$. The input tensor is convolved with all three types of filters, using padding to keep the output shape the same as the input shape. Eventually, the three outputs are summed to recover the desired number of output channels.

To visualize the quality of up-scaling achieved by the final up sampling step, we down-scale the real $256^3$ samples to $64^3$, and then provide them as inputs $I_2$ to the WGAN $M_1$.
Figure \ref{fig:256_slices} shows the middle slices of two real and fake samples $I_1$. Although, the up-sampling factor is $4$, the WGAN $M_1$ is able to produce convincing samples even in terms of high frequency components. 

\begin{figure*}[th!]
    \centering
    \vspace{0.25cm}
    \final{    
    }{
    \includegraphics[width=0.45\textwidth]{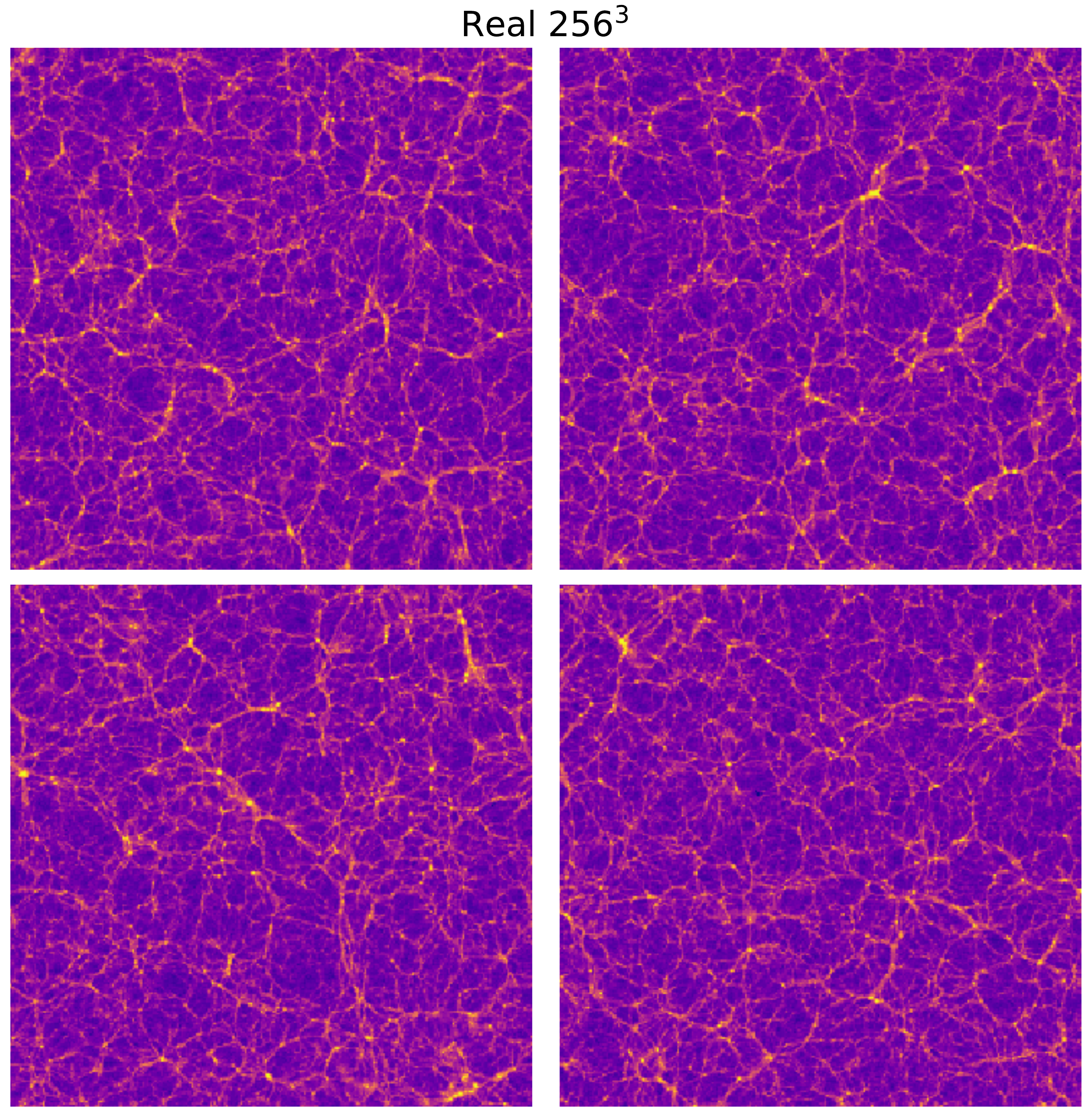}
    \hspace{0.5cm}
    \includegraphics[width=0.45\textwidth]{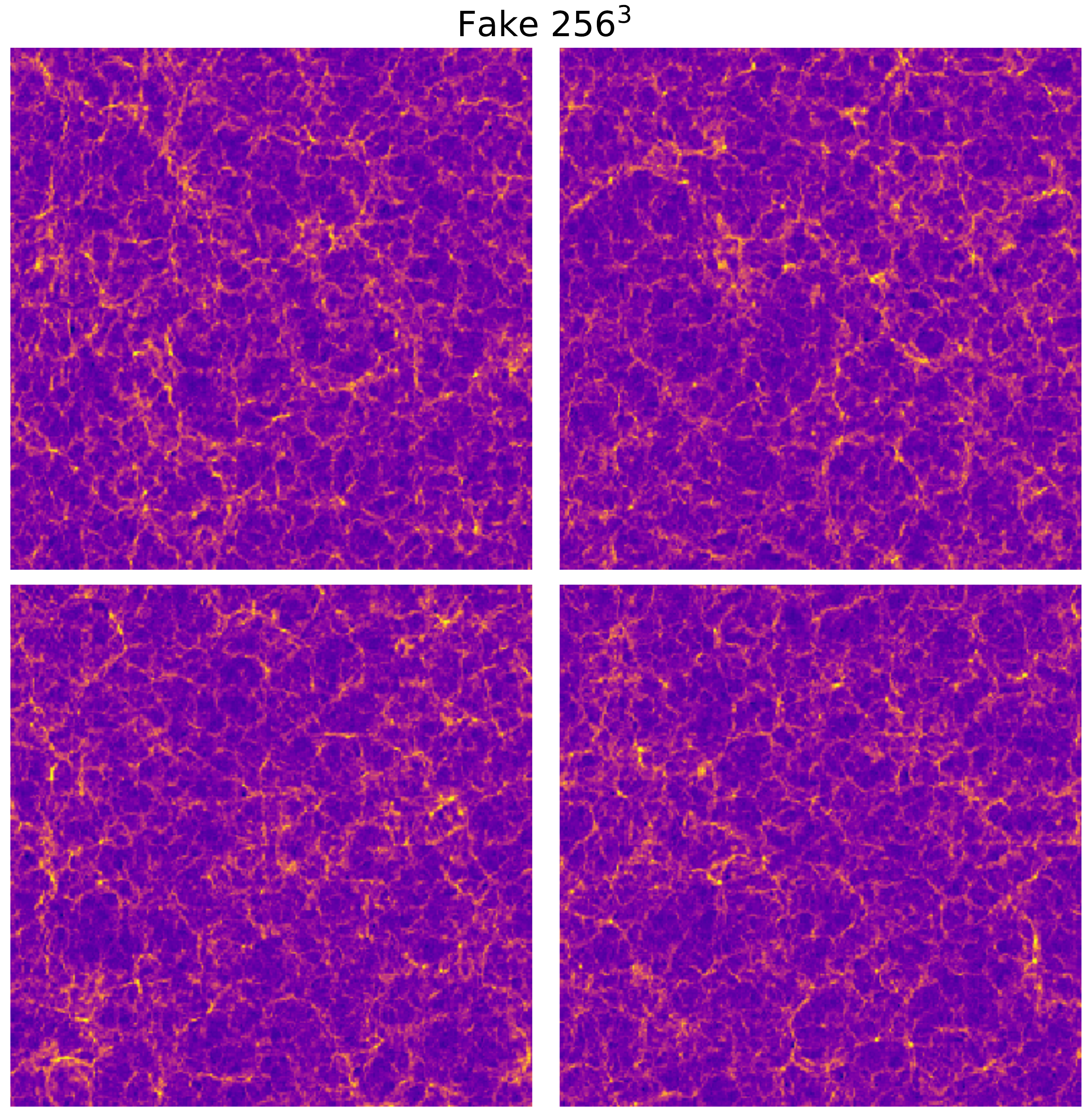}
    }
    \caption{Middle slice from real and generated $256^3$ samples. The GAN-generated samples are produced using the full multi-scale pipeline.
    Videos: \newline
     - 32-scale: \url{https://youtu.be/uLwrF73wX2w}\newline
     - 64-scale: \url{https://youtu.be/xI2cUuk3DRc}\newline
     - 256-scale: \url{https://youtu.be/nWXP6DVEalA}
    }
    \label{fig:256_full_fake}
    \centering
        \vspace{0.5cm}
    \final{    
    }{
        \includegraphics[width=0.3\textwidth]{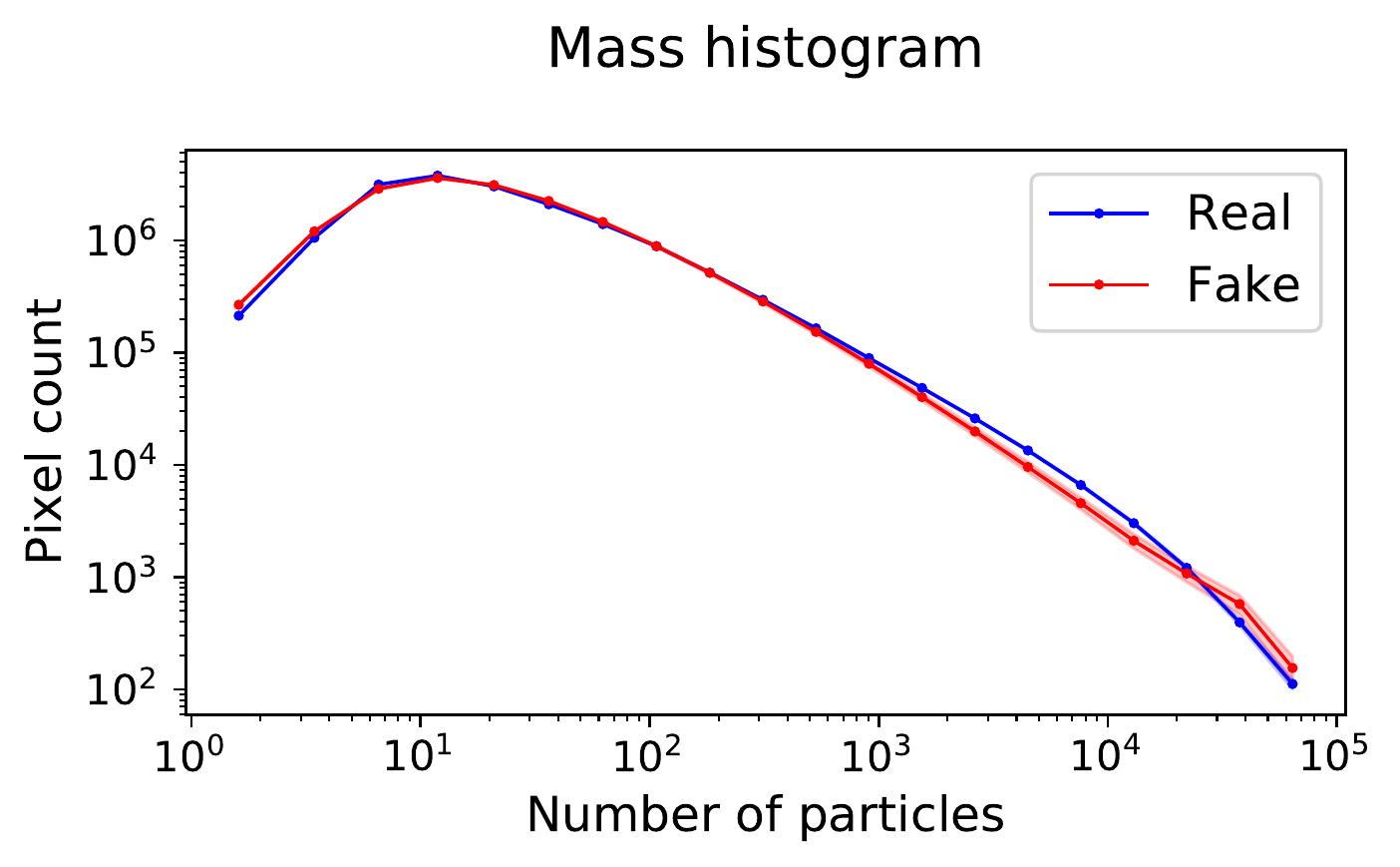}
        \includegraphics[width=0.3\textwidth]{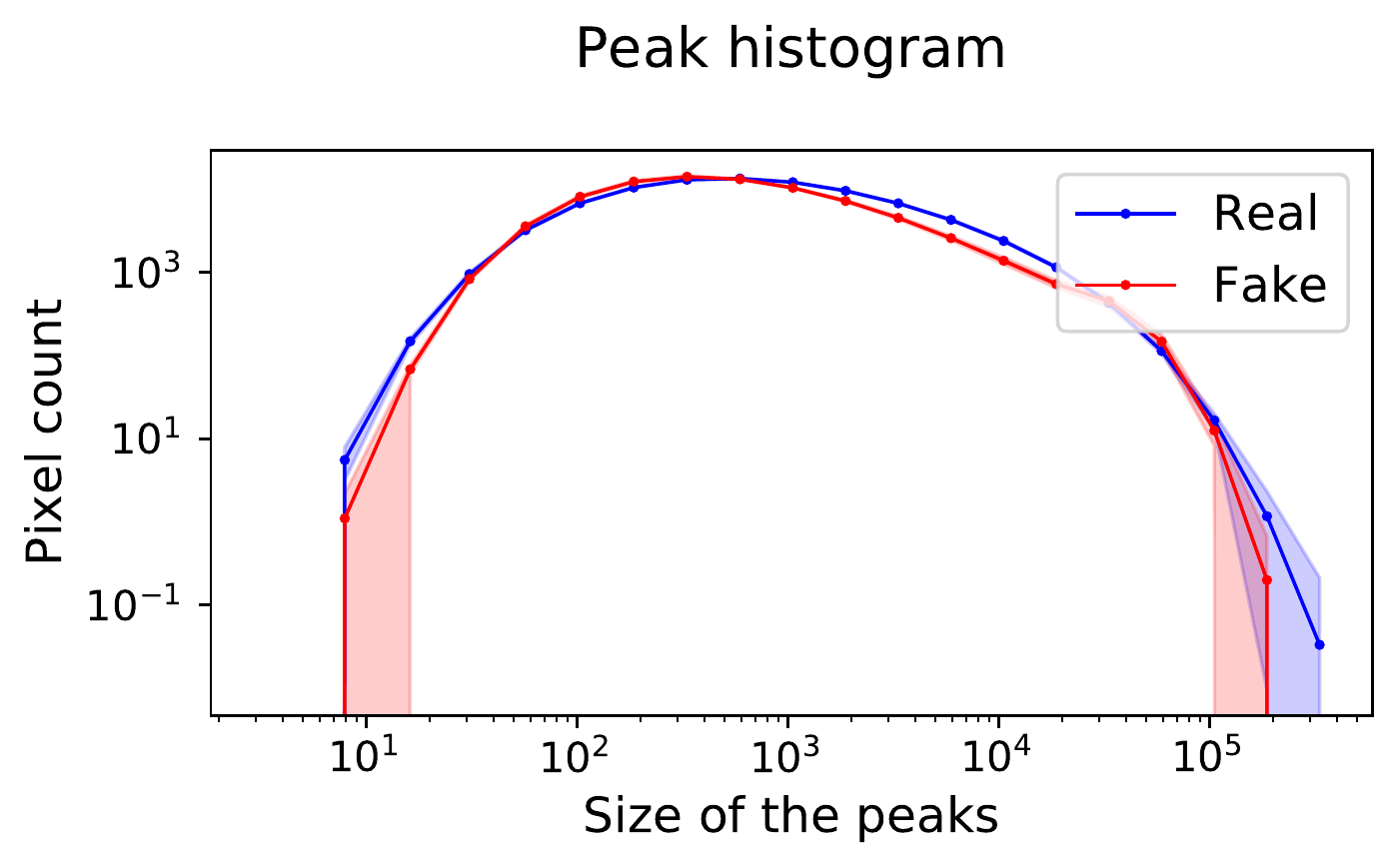}
        \includegraphics[width=0.3\textwidth]{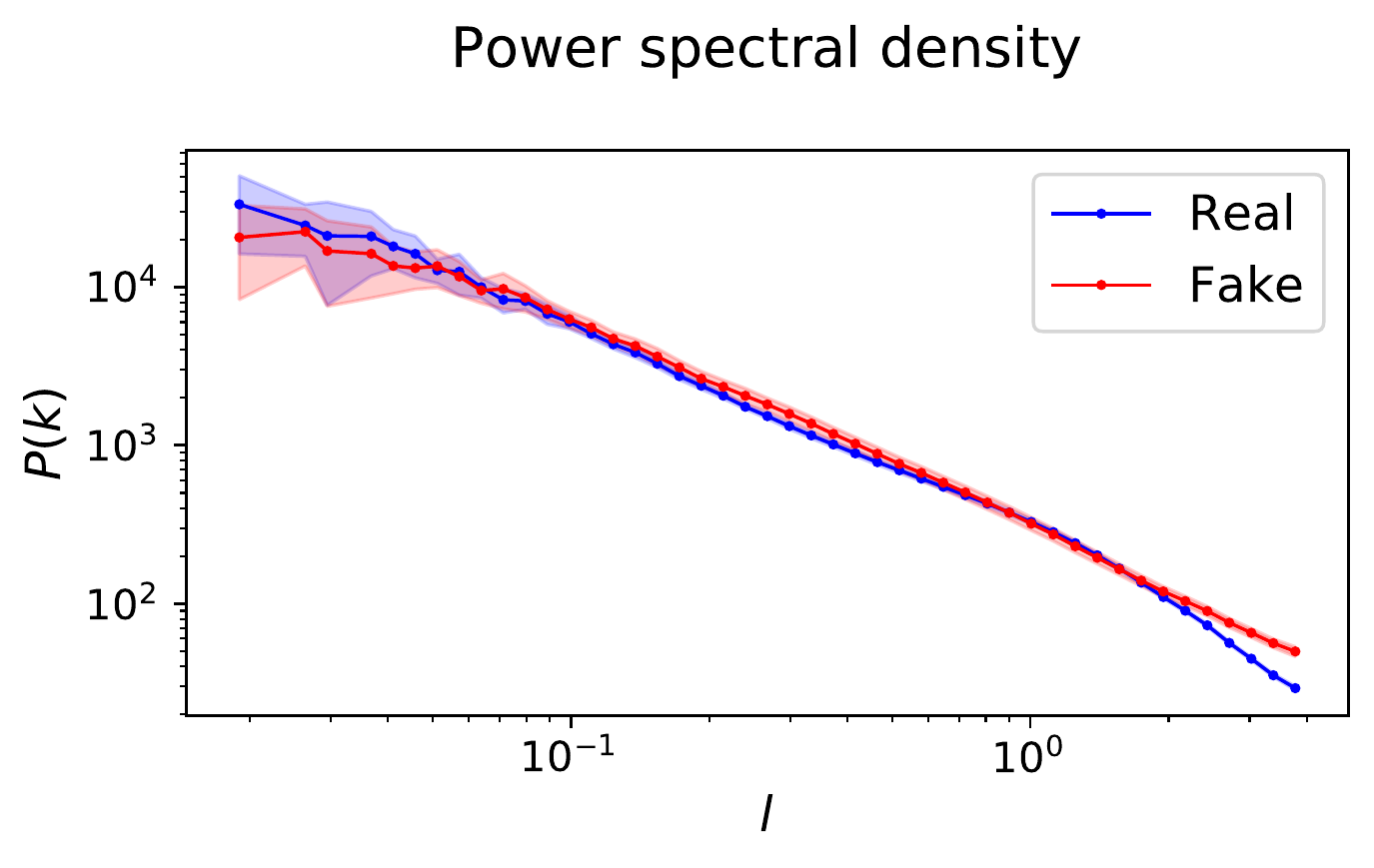}
    }
    \caption{\small{Summary statistics of real and GAN-generated $256^3$ images using the full multi-scale pipeline. The power spectrum density is shown in units of h Mpc$^{-1}$, where h = H$_0$/100 corresponds to the Hubble parameter.
    }
    \label{fig:256_full_fake_stat}
    }

    \vspace{0.5cm}
  \hrulefill
    \vspace{0.5cm}

    \begin{minipage}[l]{0.9\columnwidth}
       \centering
    \final{
    }{
    \includegraphics[width=0.9\linewidth]{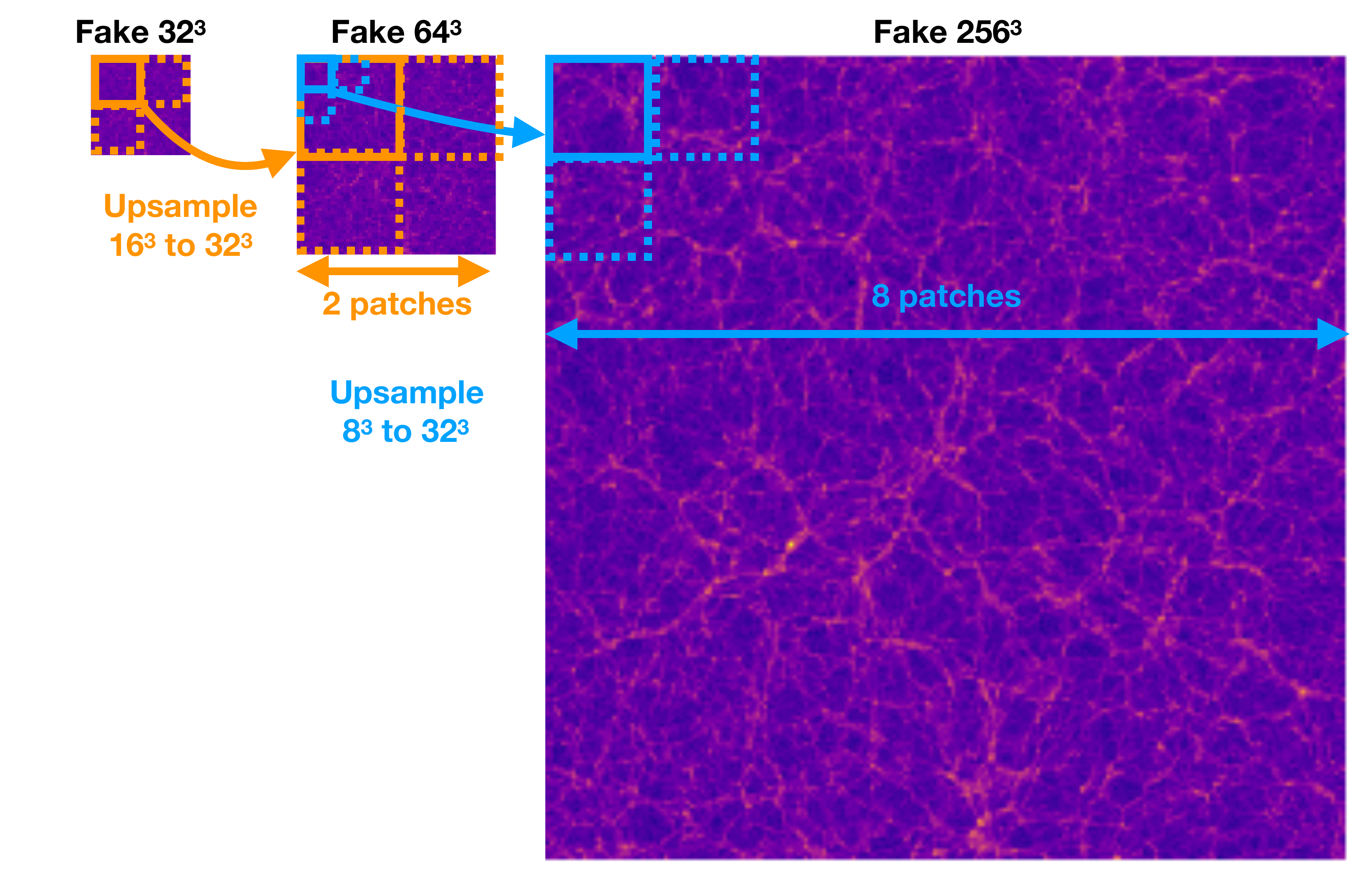}
    }
    \caption{\label{fig:Patch-Upscale}Up-scaling patches in 3-steps, using 3 different WGANs}
    \end{minipage}
    \hspace{0.7cm}
    \begin{minipage}[r]{0.9\columnwidth}
       \centering
\begin{tabular}{l|c|c|c}
Score $S^*$ & PSD & Mass Hist. & Peak Hist\\
 \hline
Multiscale            & 2.72 & 5.72 & 0.63 \\
Uniscale           & 0.01 & 0.15 & 0.07 \\
Reference   & 8.16 & 1433 & 8.17 \\
\end{tabular}
\captionof{table}{\label{tab:score-3d} Scores computed for the 30 GAN-generated $256^3$ images using the full multi-scale pipeline. We refer to \eqref{eq:score} and \eqref{eq:frechet_distance} for details.
The reference is computed by comparing two sets of 15 real samples. It represent the best score that can be achieved.}
    \end{minipage}
\end{figure*}


\subsection{Full Multi-scale Pipeline}
\label{sec:exp:full_pipeline}

\paragraph{Sample generation}
The generation process used to produce new samples proceeds as follows.
First, a latent variable is randomly drawn from the prior distribution and is fed to $M_3$ which produces a $32^3$ low-resolution cube. The latter is then upsampled by $M_2$. At first, all border patches shown in Figure~\ref{fig:Multi-Scale-Model} are set to zero. Then the $64^3$ cube is built recursively (in $2^3=8$ steps) where at each step, the previously generated patches are re-fed as borders into the generator. The generation of the full $256^3$ cube is done in a similar fashion by $M_1$. Note that this last step requires the generation of $8^3=512$ patches/smaller cubes. An illustration, adapted to the 2D case for simplicity, is shown in Figure \ref{fig:Patch-Upscale}.
The full generation process takes around $7$ seconds to produce one sample of size $256^3$ using a single GPU node with 24 cores compared to approximately 6-8 hours for a fast and approximate \textsc{L-Picola} \cite{lpicola} simulator running on two identical nodes.

\paragraph{Quantitative results}
Figure \ref{fig:256_full_fake} shows a few slices from a \mbox{3D} fake $I_1$ image generated using the full pipeline, alongside a random real $I_1$ image. Figure~\ref{fig:256_full_fake_stat} shows the summary statistics of 500 GAN generated samples, compared to that of real samples.
The visual agreement between the real and generated samples is good, although a trained human eye can still distinguish between real and fake samples.
In particular, a careful visualization reveals that the transitions between the different patches are still imperfect and that the generated samples have less long range filaments than the true samples.
The summary statistics agree very well for the middle range of mass density, with slight disagreements at the extremes.
The shape of the power spectrum density (PSD) matches well, but the overall amplitude is too high for most of the $k$ range.
Naturally, we would expect the error of the multi-scale pipeline to be the result of the accumulation of errors from the three upsampling steps. In practice, we observe a lot of similarities between the statistics shown in Figure~\ref{fig:256_stat} (from the last upscaling step) and in the Figure~\ref{fig:256_full_fake_stat} (from the full pipeline). 
Finally, Table~\ref{tab:score-3d} presents the scores obtained for the 30 cubes of size $256^3$ using the full multi-scale pipeline as generated by the full GAN sequence.
As explained in \ref{sec:score}, the reference score gives an indication of the variability within the training set, i.e, how similar are two independent sample sets from the training set. The reference mass histogram score is much higher than the PSD and the Peak histogram due to the fact that this statistic has in general much less variance. 
For that reason, it is probably easier to estimate, as indicated by the score of our pipeline.
Cosmological analyses typically require the precision of less than few percent on these summary statistics, which is achieved by the GAN method only for specific scales, peak and density values.

The scores of multiscale approach are much higher than the ones of the simpler single-scale approach described in the following section.

\begin{figure*}
    \centering
    \final{    
    }{
    \includegraphics[width=0.45\textwidth]{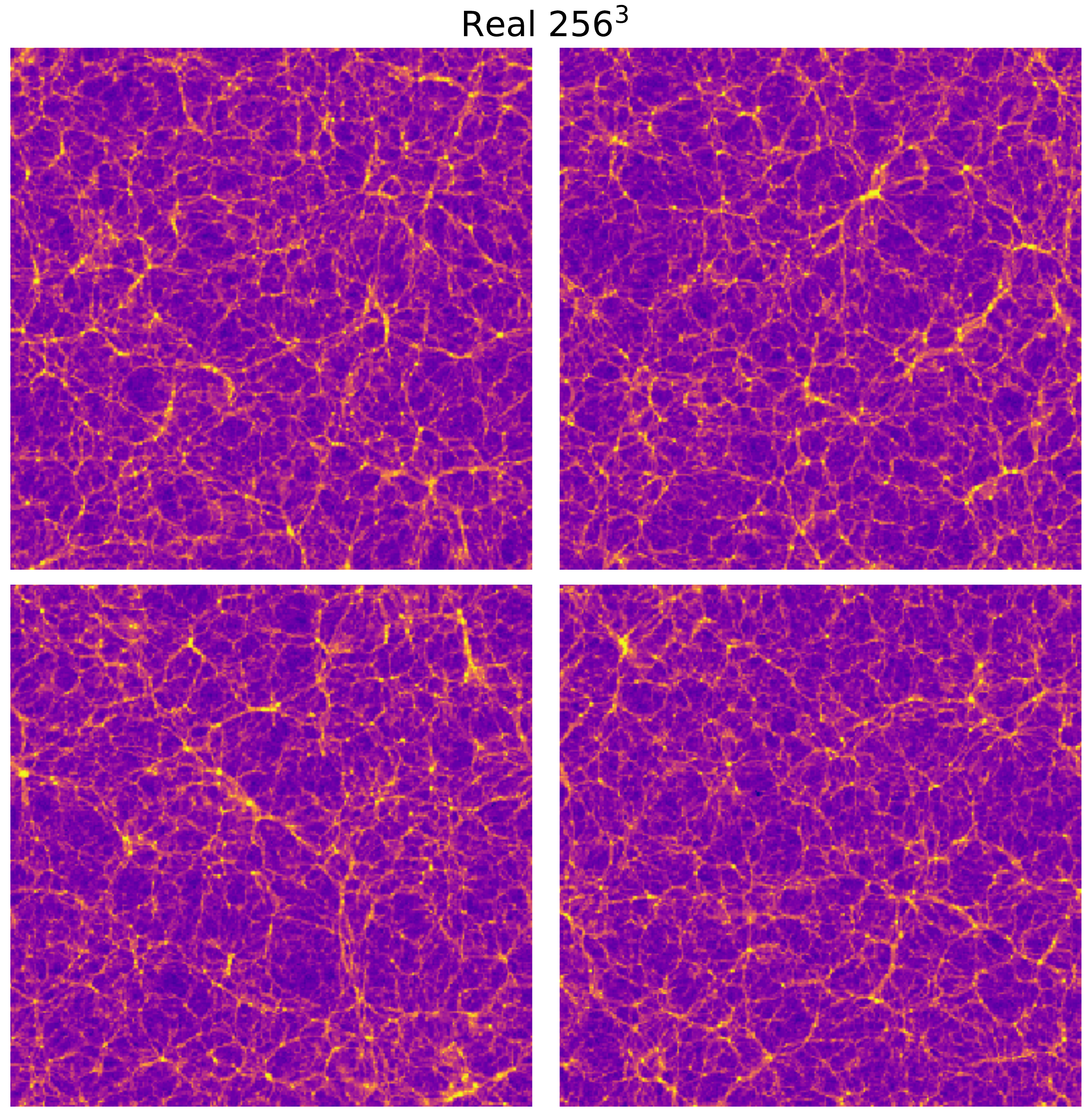}
    \hspace{0.5cm}
    \includegraphics[width=0.45\textwidth]{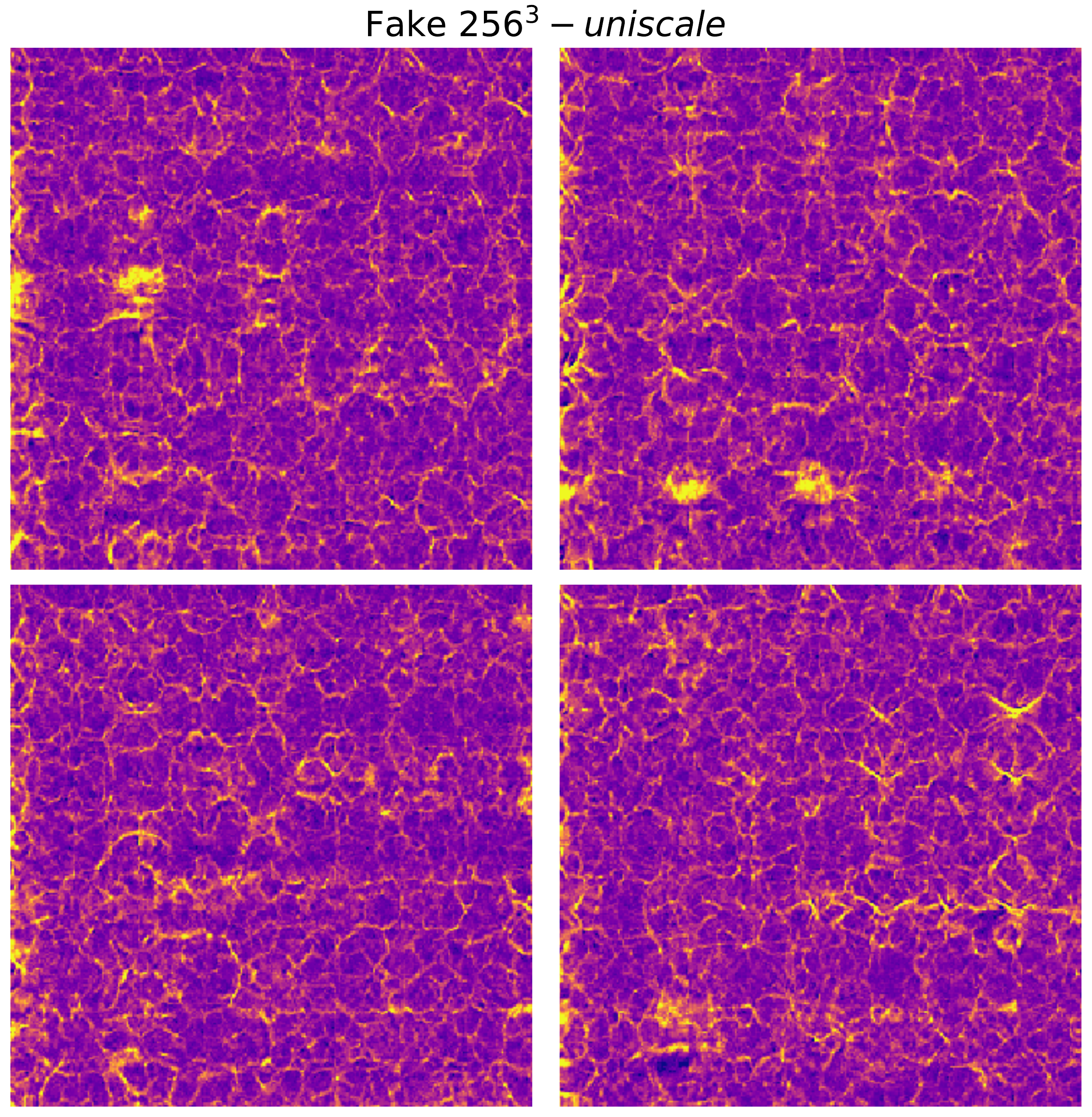}
    }
    \caption{Middle slice from real and generated $256^3$ samples. The GAN-generated samples are produced using the full uni-scale pipeline. Video: \url{https://youtu.be/fxZEQHEGunA}
    }
    \label{fig:256_uniscale_fake}
    \centering
    \final{    
    }{
        \includegraphics[width=0.3\textwidth]{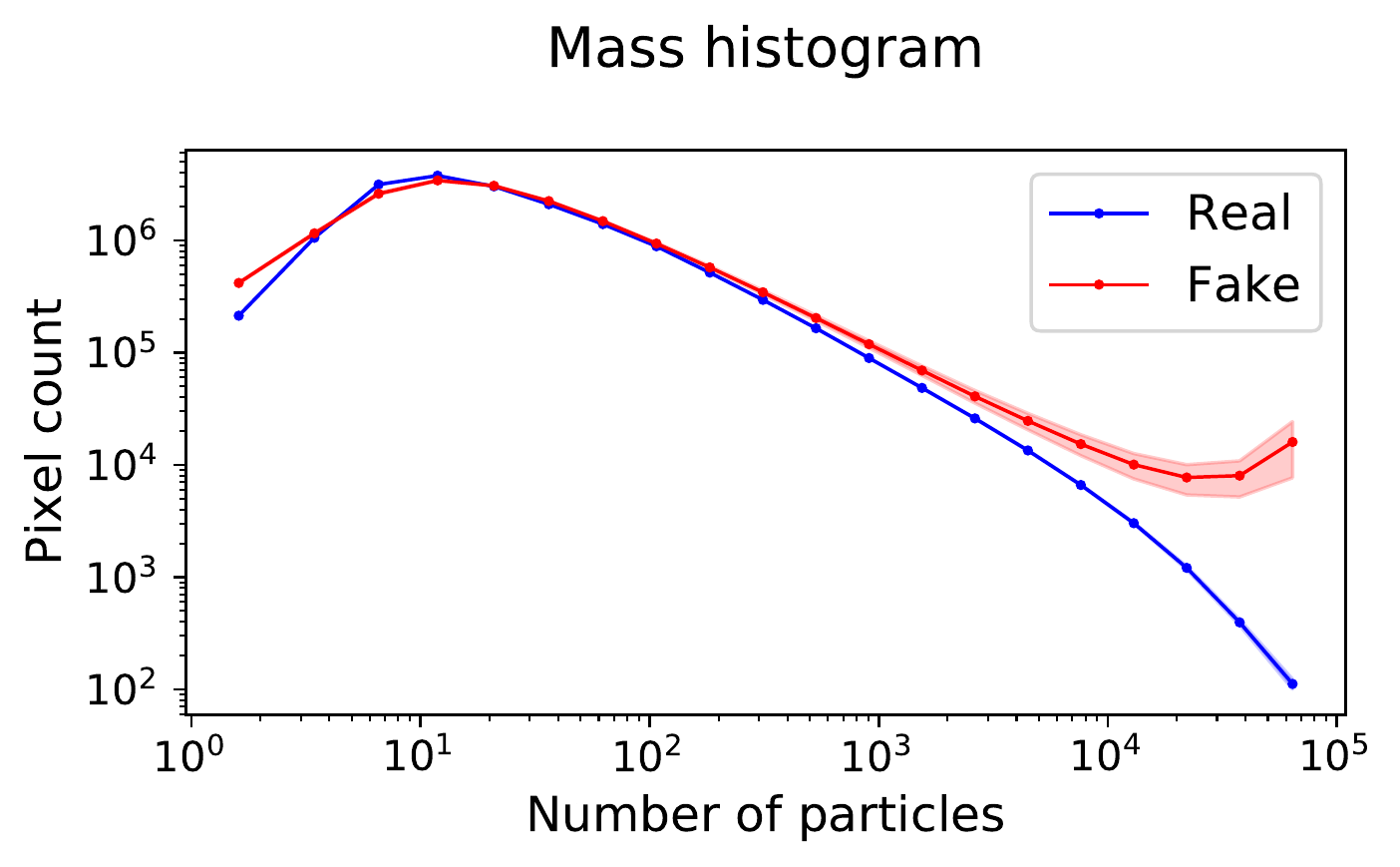}
        \includegraphics[width=0.3\textwidth]{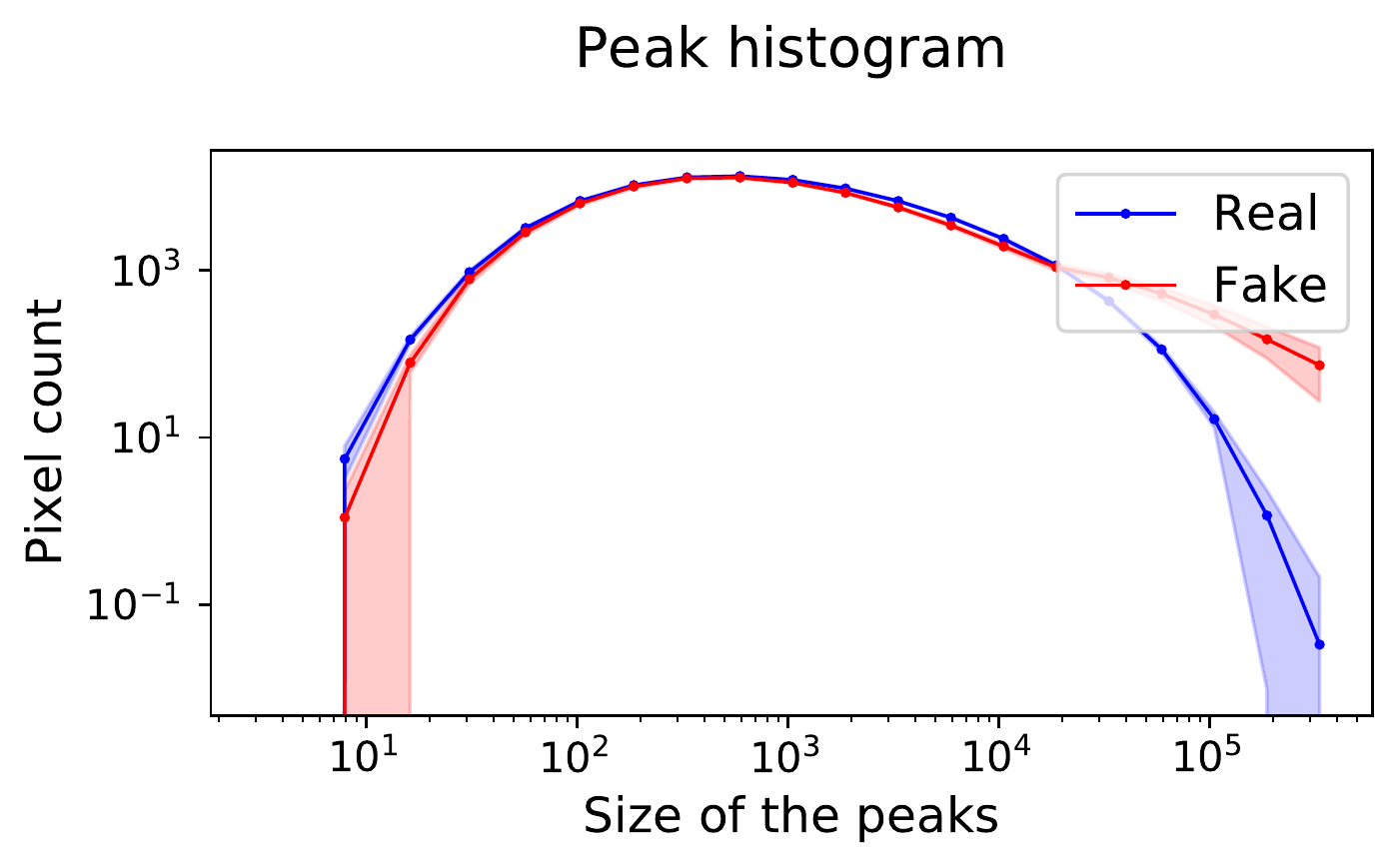}
        \includegraphics[width=0.3\textwidth]{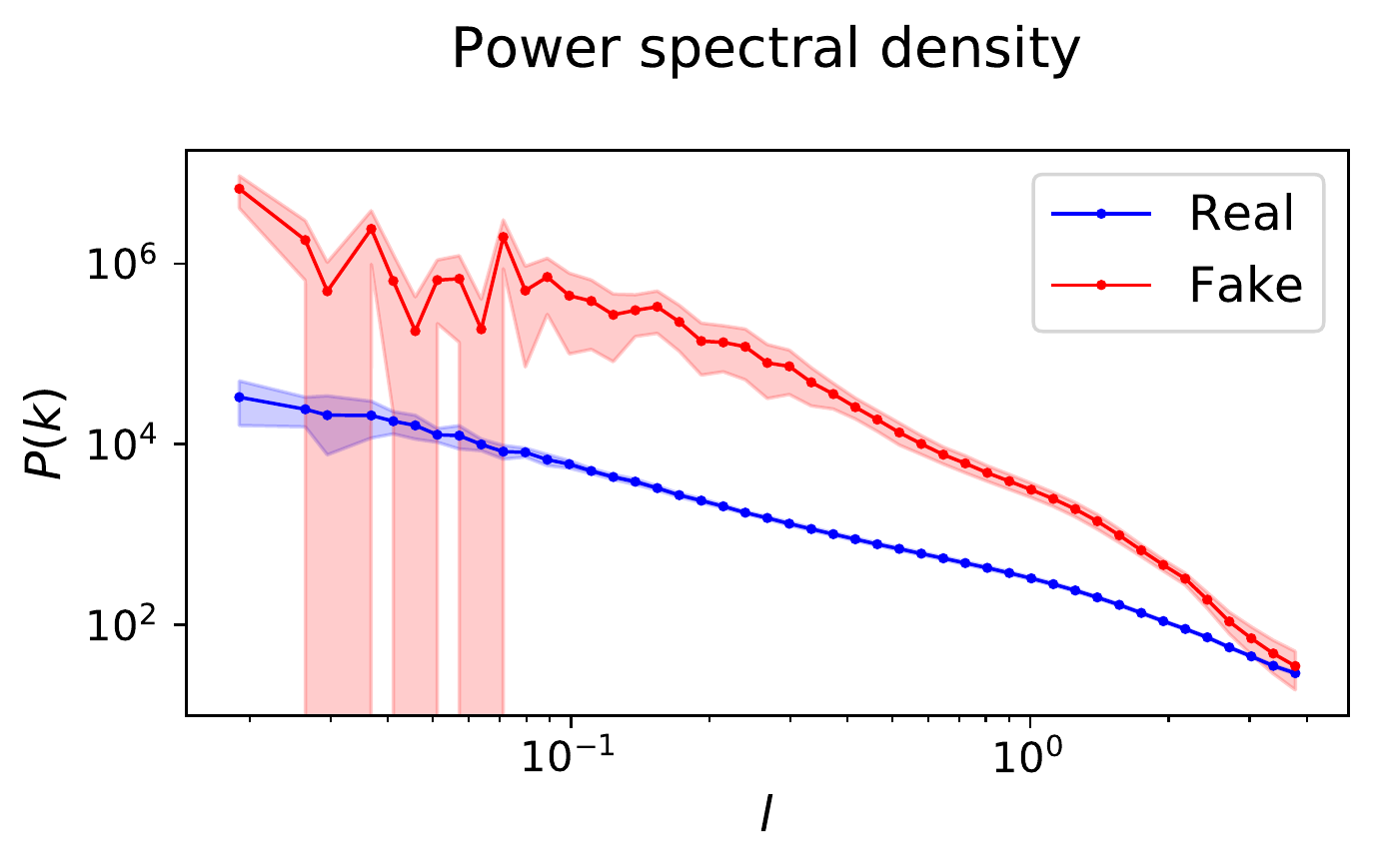}
    }
    \caption{\small{Summary statistics of real and GAN-generated $256^3$ images using the full uni-scale pipeline. The power spectrum density is shown in units of h Mpc$^{-1}$, where h = H$_0$/100 corresponds to the Hubble parameter.}
    }
    \label{fig:256_uniscale_fake_stat}
\end{figure*}

\subsection{Comparison to the single scale model}
\label{sec:single-scale-model}
In order to evaluate the effectiveness of the multi-scale approach, we compare our model to a uni-scale variant that is not conditioned on the down-sampled image. 
Here each patch is generated using only its neighboring $7$ border patches and a latent variable that is drawn from the prior distribution.
More simply, it is a direct implementation of the principle displayed in Figure \ref{fig:Multi-Scale-Model} left in 3D, where the discriminator is \emph{not} conditioned on the down-sampled image.
An important issue with the uni-scale model is that the discriminator never s more than $64^3$ pixels at once. Hence, it is likely to fail to capture long-range correlations. Practically, we observed that the training process is unstable and subject to mode collapse. At generation time. the recursive structure of the model often lead to repeating pattern.
The resulting models were of bad quality as shown in Figure \ref{fig:256_uniscale_fake} and \ref{fig:256_uniscale_fake_stat}. 
This translates to the score presented in table \ref{tab:score-3d}.
This experiment demonstrates that conditioning on  data at lower scales does play an important role in generating samples of good quality.

\section{Conclusion}

In this work we introduced a new benchmark for the generation of 3D $N$-body simulations using deep generative models.
The dataset is made publicly available and contains matter density distributions represented as cubes of $256\times256\times256$ voxels.
While the performance of the generative model can be measured by visual inspection of the generated samples, as commonly done on datasets of natural images, we also offer a more principled alternative based on a number of summary statistics that are commonly used in cosmology.
Our investigation into this problem has revealed that several factors make this task challenging, including:
(i) the sheer volume of each data sample, which is not straightforwardly tractable using conventional GAN architectures,
(ii) the large dynamic range of the data that spans several orders of magnitude; which requires a custom-designed transformation of the voxel values, and
(iii) the need for high accuracy required for the model to be practically usable for a real cosmological study. Adding to the difficulties of (i) and (ii), this also requires accurately capturing features that are rare in the training set.

As a first baseline result for the newly introduced benchmark, we proposed a new method to train a deep generative model on \mbox{3D} images.
We split the generation process into the generation of smaller patches as well as condition on neighboring patches.
We also apply a multi-scale approach to learn multiple WGANs at different image resolutions, each capturing salient features at different scales.
This approach is inspired by Laplacian pyramid GAN \cite{LAP_GAN} and by PixelCNN \cite{Pixel_CNN}, which have both been developed for 2D data.

We find that the proposed baseline produces $N$-body cubes with good visual quality compared to the training data, but significant differences can still be perceived.
Overall, the summary statistics between real and generated data match well, but notable differences are present for high voxel values in the mass and peak histograms.
The power spectrum has the expected shape, with amplitude that is too high for most of the $k$ values.
The overall level of agreement is promising, but can not yet be considered as sufficient for practical applications in cosmology.
Further development will be needed to achieve this goal; in order to encourage it, we have made the dataset and the code publicly available.

In our current model, the discriminator only has access to a partial view of the final image.
The dependencies at small scale that may exist between distant patches are therefore not captured by the discriminator.
Extending this model to allow the discriminator to have a more global view would be the next logical extension of this work. We have also observed empirically that the extreme right tail of the histogram is often not fully captured by the generator. Designing architectures that would help the generative model to handle large dynamic range in the data could further improve performance.
One could also get further inspiration from the literature on generative models for video data, such as~\cite{vondrick2016generating,Xiong_2018_CVPR,saito2017temporal}. Given the observations made in our experiments, one might for instance expect that the two stage approach suggested in~\cite{saito2017temporal} could address some of the problem seen with the right tail of the distribution.

Another interesting research direction would be to condition the generation process on cosmic time or cosmological parameters. One could for instance rely on a conditional GAN model such as \cite{Mirza2014conditionalgans}.

\appendix

\section{Input Data Transformation}
\label{sec:input_transform}
The general practice in machine learning is that the input data is first standardized before giving it to any machine learning model.
This preprocessing step is important as many neural networks and optimization blocks are scale-dependent, resulting is most architectures working optimally only when the data is appropriately scaled.
Problematically, because of the physical law of gravity, most of the universe is empty, while most of the matter is concentrated in a few small areas and filaments.
The dataset had the minimum value of $0$ and the maximum value of $185,874$, with most of the voxels concentrated close to zero, and significantly skewed towards the smaller values and has an elongated tail towards the larger ones.
Even with standardization, it is difficult for a generative model to learn very sparse and skewed distributions.
Hence we transform the data using a special function, in a slightly different way than \cite{rodriguez2018fast}.

In order to preserve the sensitivity of the smaller values, a logarithm-based transformation function is a good candidate.
Nevertheless, to maintain the sensitivity to the large values, we should favor a linear function.
In our attempt to coincide the best of the two regimes, we design a function that is logarithmic for lower values and linear for large one, i.e after a specified cutoff value.
The exact forward transformation function, $y = f(x, c, s)$
is defined as:
\begin{equation}
 f(x, c, s)  =  f^{\prime}(x+s, c) - f^{\prime}(s, c),
 \end{equation}
where
\begin{equation}
  f^{\prime}(x, c) =
    \begin{cases}
      3\log_e (x+1) & \text{if $x \leq c$}\\
      3(\log_e (c+1) + \frac{x}{c} - 1) & \text{otherwise.}\\
    \end{cases}
\end{equation}
As a result, the backward transformation function $x = b(y, c, s)$ reads
\begin{equation}
b(y, c, s) = b^{\prime}( y  + f^{\prime}(s, c),\ c) - s,
\end{equation}
where
\begin{equation}
  b^{\prime}(y, c) =
    \begin{cases}
      e^{\frac{y}{3}} - 1 & \text{if $y \leq 3\log_e(c+1)$}\\
      c(\frac{y}{3} + 1 - \log_e(c+1)) & \text{otherwise.}\\
    \end{cases}
\end{equation}
Here $c$ and $s$ are selected hyper-parameters.
For our experiments we found $c=20,000$ and $s=3$ to be good candidates.
After the forward transformation, the distribution of the data becomes similar to a one-sided Laplacian distribution.
We always invert this transformation once new data is generated, before calculating the summary statistics.

\section{$N$-body Training Set Augmentation}
\label{sec:augmentation}
As $N$-body simulations are very expensive, we need to make sure to use all the information available using an optimal dataset augmentation.
To augment the training set, the cubes are randomly rotated by multiples of 90 degrees and randomly translated along one of the 3 axes.
The cosmological principle states that the spatial distribution of matter in the universe is homogeneous and isotropic when viewed on a large enough scale. As a consequence there should be no observable irregularities in the large scale structure over the course of evolution of the matter field that was initially laid down by the Big Bang \cite{Dodelson2003cosmology}. Hence, the rotational and translational augmentations do not alter the data distribution that we are trying to model in any way.
Moreover, we note that use circular translation in our augmentation scheme. This is possible because $N$-body simulations are created using the {\it periodic boundary condition}: a particle exiting the box on one side enters it immediately on the opposite side. Forces follow the same principle.
This prevents the particles from collapsing to the middle of the box under gravity.
These augmentations are important given that we only have 30 $N$-body cubes in our training set.


\section{Architecture \& implementation details}
\label{sec:architecture_details}

\paragraph{Implementation details}
We used Python and Tensorflow to code the models which are trained on GPUs with 16GB of memory. 
All the GANs are WGANs with a Gaussian prior distribution.
Using a single GPU, it takes around $7$ seconds to produce one sample of size $256^3$ compared to approximately 30 hours for a precise N-body simulator running on two nodes with 24 cores and a GPU, such as \textsc{PkdGrav3} \citep{Sim_2}. In this project, a  fast and approximate \textsc{L-Picola} \citep{lpicola} simulator was used, with approximately 6 hours of runtime on two nodes.

The batch size is set to 8 for all the experiments. 
All Wasserstein GANs were trained with using a gradient penalty loss with $\gamma_{GP}=10$, as described in \cite{gulrajani2017improved}.
We use RMSprop with a learning rate $3\cdot10^{-5}$ for both the generator and discriminator. The discriminator was updated $5$ times per generator update.

\paragraph{Network architectures}
The neural networks used in our experiments are variants of deep convolutional networks with inception modules and/or residual connections.

All weights were initialized using Xavier Initializer, except the bias that was initialized to $0$.
We used leaky ReLu and spectral normalization \cite{spectral_norm} to stabilize the network.
The architectures are detailed in Tables~\ref{tab:archi32},~\ref{tab:archi64} and~\ref{tab:archi256}.

Handling the input border is an architectural challenge in itself. We used two different solutions to overcome this issue and use one of them for each scale.

\paragraph{Generator $M_2$.}
The generator $M_2$ possesses a convolutional encoder for the borders. Once the borders are encoded, we concatenate them with the latent variable. The downsampled image simply concatenated at the first convolution layer (see Table~\ref{tab:archi64}).

\paragraph{Generator $M_1$.}
The generator $M_1$ does not possess an encoder, but utilize the border directly as extra channels. As a result, the generator convolution all have a stride of $1$. The downsampled image is first upsampled using a simple transposed convolution with a constant kernel and then concatenated as an input. The latent variable is of size $32^3$ to avoid a memory consuming linear layer.
Eventually, as the convolution is shift-invariant, we perform two transformations to the input borders before feeding them to the generator. As a results, we flip them to obtain a correct alignment with produced corner.
Furthermore, to improve the capacity of the networks without increasing to much the number of parameters and channels, we use an inception inspired module. The module is simply composed of 3 convolutions ($1\times 1\times 1$, $2\times 2\times 2$, $4 \times 4 \times4$) in parallel followed by a merging $1 \times 1 \times 1$ convolution.
Finally, to further help the discriminator, we also feed some PSD estimation at the beginning of its linear layer (see Table~\ref{tab:archi256}).

\paragraph{Training stabilization using a regularizer.}
While it has been shown that the gradient penalty loss of the Wasserstein GAN helps in stabilizing the training process~\cite{gulrajani2017improved}, this term does not prevent the discriminator to saturate. For example, when the discriminator has a high final bias, its output will be very large for both real and fake sample, yet its loss might be controlled as the output of real samples is subtracted from the one of the fake samples. In practice, we noticed that when this behavior was happening, the learning process of the generator was hindered and the produced samples were of worse quality.
In order to circumvent this issue, we added a second regularization term:
\begin{equation}
\text{ReLu}(D_{real} \cdot D_{fake})
\end{equation}
Our idea was that the regularization should kick in only to prevent the un-desirable effect and should not affect the rest of the training. If the discriminator is doing a good job, then $D_{real}$ should be positive and $D_{fake}$ negative nullifying the regularization. On the contrary if both of these term are of the same sign, the output will be penalized quadratically forcing it to remain close to $0$. While the effect of this second regularization term is still unclear to us, it did help to stabilize our optimization procedure for the multi-scale approach.

As we release our code and entire pipeline, we encourage the reader to check it for additional details.

\begin{table}[h]
\centering
\footnotesize
\begin{tabular}{|l|l|l|}
 \hline
  Operation & Parameter size& Output Shape  \\
 \hline
 \multicolumn{3}{c}{\vspace{-0.1cm}} \\
\multicolumn{3}{c}{Generator} \\
\hline
Input $z ~ \mathcal{N}(0,1)$ & & $(n,256)$\\
Dense & $(256, 256d)$ & $(n, 256d)$\\
Reshape & & $(n,4,4,4,4d)$ \\
TrConv 3D (Sride 2) & $(4,4,4,4d,4d)$ & $(n, 8, 8, 8, 4d)$\\
LReLu $(\alpha=0.2)$  &  & $(n, 16, 16, 16, 4d)$\\
TrConv 3D (Sride 2) & $(4,4,4,4d,2d)$ & $(n, 16, 16, 16, 2d)$\\
LReLu $(\alpha=0.2)$  &  & $(n, 16, 16, 16, 2d)$\\
TrConv 3D (Sride 2) & $(4,4,4,d,d)$ & $(n, 32, 32, 32, d)$\\
LReLu $(\alpha=0.2)$  &  & $(n, 32, 32, 32, 2d)$\\
TrConv 3D (Sride 1) & $(4,4,4,d,d)$ & $(n, 32, 32, 32, d)$\\
LReLu $(\alpha=0.2)$  &  & $(n, 32, 32, 32, 2d)$\\
TrConv 3D (Sride 1) & $(4,4,4,d,1)$ & $(n, 32, 32, 32, 1)$\\
\hline
\multicolumn{3}{c}{\vspace{-0.1cm}} \\
\multicolumn{3}{c}{Discriminator} \\
\hline
Input generated image & & $(n, 32, 32, 32, 1)$\\
Conv 3D (Sride 2) & $(4,4,4,1,d) $& $(n, 32, 32, 32, d)$\\
LReLu $(\alpha=0.2)$  &  & $(n, 32, 32, 32, d)$\\
Conv 3D (Sride 2) & $(4,4,4,d,d)$ & $(n, 32, 32, 32, d)$\\
LReLu $(\alpha=0.2)$  &  & $(n, 32, 32, 32, d)$\\
Conv 3D (Sride 1) & $(4,4,4,d,2d)$ & $(n, 16, 16, 16, 2d)$\\
LReLu $(\alpha=0.2)$  &  & $(n, 16, 16, 16, 2d)$\\
Conv 3D (Sride 1) & $(4,4,4,2d,4d)$ & $(n, 8, 8, 8, 4d)$\\
LReLu $(\alpha=0.2)$  &  & $(n, 8, 8, 8, 4d)$\\
Conv 3D (Sride 1) & $(4,4,4,4d,8d)$ & $(n, 4, 4, 4, 8d)$\\
LReLu $(\alpha=0.2)$  &  & $(n, 4, 4, 4, 8d)$\\
Reshape &  & $(n, 512d)$\\
Dense & $(512d, 1)$ & $(n,1)$ \\
\hline
\end{tabular}
\caption{Detailled architecture of the low resolution GAN $0\rightarrow 32^3$. $d=64$.
\label{tab:archi32}}
\end{table}

\begin{table}[h]
\footnotesize
\centering
\begin{tabular}{|l|l|l|}
 \hline
  Operation & Parameter size& Output Shape  \\
 \hline
 \multicolumn{3}{c}{\vspace{-0.1cm}} \\
\multicolumn{3}{c}{Generator} \\
\hline
Input borders & & $(n, 32, 32, 32, 7)$\\
Conv 3D (Sride 2) & $(4,4,4,7,d)$ & $(n, 16, 16, 16, d)$\\
Conv 3D (Sride 2) & $(4,4,4,d,d)$ & $(n, 8, 8, 8, d)$\\
Conv 3D (Sride 2) & $(4,4,4,d,16)$ & $(n, 4, 4, 4, 16)$\\
Reshape &  & $(n, 1024)$\\
Input $z_1 ~ \mathcal{N}(0,1)$ & & $(n,1024)$\\
Concatenation & & $(n,2048)$\\
Dense & $(2048, 256d)$ & $(n, 256d)$\\
Reshape & & $(n,16,16,16,2)$ \\
Input downsampled corner &  & $(n,16,16,16,1)$\\
Input $z_2 ~ \mathcal{N}(0,1)$ & & $(n,16,16,16,1)$\\
Concatention & & $(n,16,16,16,4)$\\
TrConv 3D (Sride 1) & $(4,4,4,4,d)$ & $(n, 16, 16, 16, d)$\\
LReLu $(\alpha=0.2)$  &  & $(n, 16, 16, 16, 4d)$\\
TrConv 3D (Sride 1) & $(4,4,4,d,d)$ & $(n, 16, 16, 16, d)$\\
LReLu $(\alpha=0.2)$  &  & $(n, 16, 16, 16, 2d)$\\
TrConv 3D (Sride 2) & $(4,4,4,d,4d)$ & $(n, 32, 32, 32, 4d)$\\
LReLu $(\alpha=0.2)$  &  & $(n, 32, 32, 32, 4d)$\\
TrConv 3D (Sride 1) & $(4,4,4,4d,2d)$ & $(n, 32, 32, 32, 2d)$\\
LReLu $(\alpha=0.2)$  &  & $(n, 32, 32, 32, d)$\\
TrConv 3D (Sride 1) & $(4,4,4,2d,1)$ & $(n, 32, 32, 32, 1)$\\
\hline
\multicolumn{3}{c}{\vspace{-0.1cm}} \\
\multicolumn{3}{c}{Discriminator} \\
\hline
Input generated image & & $(n, 32, 32, 32, 1)$\\
Input borders & & $(n, 32, 32, 32, 7)$\\
Reshape to a cube & & $(n, 64, 64, 64, 1)$\\
Input smooth image & & $(n, 64, 64, 64, 1)$\\
Concatenation (+ diff) & & $(n, 64, 64, 64, 3)$\\
Conv 3D (Sride 1) & $(4,4,4,3,d)$ & $(n, 64, 64, 64, d)$\\
LReLu $(\alpha=0.2)$  &  & $(n, 64, 64, 64, d)$\\
Conv 3D (Sride 2) & $(4,4,4,d,2d)$ & $(n, 32, 32, 32, 2d)$\\
LReLu $(\alpha=0.2)$  &  & $(n, 32, 32, 32, d)$\\
Conv 3D (Sride 2) & $(4,4,4,2d,4d)$ & $(n, 16, 16, 16, 4d)$\\
LReLu $(\alpha=0.2)$  &  & $(n, 16, 16, 16, d)$\\
Conv 3D (Sride 1) & $(4,4,4,d,d)$ & $(n, 16, 16, 16, d)$\\
LReLu $(\alpha=0.2)$  &  & $(n, 16, 16, 16, d)$\\
Conv 3D (Sride 2) & $(4,4,4,d,d)$ & $(n, 8, 8, 8, d)$\\
LReLu $(\alpha=0.2)$  &  & $(n, 8, 8, 8, d)$\\
Reshape &  & $(n, 512d)$\\
Dense & $(512d, 1)$ & $(n,1)$ \\
\hline
\end{tabular}
\caption{Detailled architecture of UpscaleGAN $32^3\rightarrow 64^3$. $d=32$.
\label{tab:archi64}}
\end{table}

\begin{table}[ht]
\centering
\footnotesize
\begin{tabular}{|l|l|l|}
 \hline
  Operation & Parameter size& Output Shape  \\
 \hline
 \multicolumn{3}{c}{\vspace{-0.1cm}} \\
\multicolumn{3}{c}{Generator} \\
\hline
Input $z ~ \mathcal{N}(0,1)$ & - & $(n,32,32,32,1)$\\
Input smooth image  & - & $(n,32,32,32,1)$\\
Input borders & - & $(n,32,32,32,7)$\\
Concatention & - & $(n,32,32,32,9)$\\
InConv 3D (Sride 1) & * & $(n, 32, 32, 32, d)$\\
LReLu $(\alpha=0.2)$  &  & $(n, 32, 32, 32, d)$\\
InConv 3D (Sride 1) & * & $(n, 32, 32, 32, d)$\\
LReLu $(\alpha=0.2)$  &  & $(n, 32, 32, 32, d)$\\
InConv 3D (Sride 1) & * & $(n, 32, 32, 32, d)$\\
LReLu $(\alpha=0.2)$  &  & $(n, 32, 32, 32, d)$\\
InConv 3D (Sride 1) & * & $(n, 32, 32, 32, d)$\\
LReLu $(\alpha=0.2)$  &  & $(n, 32, 32, 32, d)$\\
InConv 3D (Sride 1) & * & $(n, 32, 32, 32, d)$\\
LReLu $(\alpha=0.2)$  &  & $(n, 32, 32, 32, d)$\\
InConv 3D (Sride 1) & * & $(n, 32, 32, 32, d)$\\
LReLu $(\alpha=0.2)$  &  & $(n, 32, 32, 32, d)$\\
InConv 3D (Sride 1) & * & $(n, 32, 32, 32, d)$\\
LReLu $(\alpha=0.2)$  &  & $(n, 32, 32, 32, d)$\\
InConv 3D (Sride 1) & * & $(n, 32, 32, 32, 1)$\\
ReLu  & - & $(n, 32, 32, 32, 1)$\\
\hline
\multicolumn{3}{c}{\vspace{-0.1cm}} \\
\multicolumn{3}{c}{Discriminator} \\
\hline
Input generated image & - & $(n, 32, 32, 32, 1)$\\
Input borders & - & $(n, 32, 32, 32, 7)$\\
Reshape to a cube & - & $(n, 64, 64, 64, 1)$\\
Input smooth image & & $(n, 64, 64, 64, 1)$\\
Concatenation (+ diff) & & $(n, 64, 64, 64, 3)$\\
InConv 3D (Sride 2) & * & $(n, 32, 32, 32, 2d)$\\
LReLu $(\alpha=0.2)$  &  & $(n, 32, 32, 32, 2d)$\\
InConv 3D (Sride 1) & * & $(n, 32, 32, 32, 2d)$\\
LReLu $(\alpha=0.2)$  &  & $(n, 32, 32, 32, 2d)$\\
InConv 3D (Sride 1) & * & $(n, 32, 32, 32, d)$\\
LReLu $(\alpha=0.2)$  &  & $(n, 32, 32, 32, d)$\\
InConv 3D (Sride 1) & * & $(n, 32, 32, 32, d)$\\
LReLu $(\alpha=0.2)$  &  & $(n, 32, 32, 32, d)$\\
InConv 3D (Sride 1) & * & $(n, 32, 32, 32, d)$\\
LReLu $(\alpha=0.2)$  &  & $(n, 32, 32, 32, d)$\\
InConv 3D (Sride 1) & * & $(n, 32, 32, 32, d)$\\
LReLu $(\alpha=0.2)$  &  & $(n, 32, 32, 32, d)$\\
InConv 3D (Sride 2) & * & $(n, 16, 16, 16, d)$\\
LReLu $(\alpha=0.2)$  &  & $(n, 16, 16, 16, d)$\\
InConv 3D (Sride 2) & * & $(n, 8, 8, 8, d)$\\
LReLu $(\alpha=0.2)$  &  & $(n, 8, 8, 8, d)$\\
Reshape & - & $(n, 16384)$\\
Compute PSD & - & $(n, 1914)$\\
Concatenate & - & $(n, 18298)$\\
Dense & $(18298, 64)$ & $(n,64)$ \\
LReLu $(\alpha=0.2)$  &  & $(n, 64)$\\
Dense & $(64 16)$ & $(n,16)$ \\
LReLu $(\alpha=0.2)$  &  & $(n, 16)$\\
Dense & $(16, 1)$ & $(n,1)$ \\
\hline
\end{tabular}
\caption{Detailed architecture of UpscaleGAN $64^3\rightarrow 256^3$. $d=64$. The parameter shape of the inception convolution written InConv is too large be written in the table.
\label{tab:archi256}}
\end{table}

\begin{backmatter}

\section*{Abbreviations}
    \begin{itemize}
        \item GAN: Generative adversarial networks
        \item WGAN: Wasserstein Generative adversarial networks
        \item DCNN: Deep convolutional neural networks
    \item LSST: Large Synoptic Survey Telescope
    \item CDM: Cold Dark Matter
    \item GPU: Graphics Processing Unit
    \item PSD: Power Spectral Density
    \item FD: Fréchet Distance
    \item KL: Kullback-Leibler
    \end{itemize}

\section*{Declarations}

\subsection*{Availability of data and material}
The data and code to reproduce the experiment of this study are available at \url{https://github.com/nperraud/3DcosmoGAN} and \url{https://zenodo.org/record/1464832}.

\subsection*{Competing interests}
The authors declare that they have no competing interests.

\subsection*{Funding}
This work was supported by the Swiss Data Science Centre (SDSC), project \textit{sd01 - DLOC:  Deep Learning for Observational Cosmology}, and grant number 200021\_169130 and PZ00P2\_161363 from the Swiss National Science Foundation.
The funding bodies had no involvement in the design of the study, collection, analysis, and interpretation of data, or writing the manuscript.

\subsection*{Author's contributions}
NP performed the experiment design and the full analysis.
NP and AS contributed to the implementation of the algorithms.
Initial implementation of the algorithm was presented in Master's thesis by AS titled ``Scalable Generative Models For Cosmology'', supervised by NP, AL, and TK, with advice from TH and AR.
NP created the challenge.
TK and AL initiated the study as the Principle Investigators of the DLOC program at the SDSC, provided the resources used in the analysis, performed the experiment design, and provided direct guidance and supervision.
JF and RS prepared the $N$-body simulation dataset.
AR and TH contributed to the development of the ideas and the proposal.
All authors read and approved the final manuscript.

\subsection*{Acknowledgments}
We thank Adam Amara for initial discussions on the project idea.
We thank Janis Fluri and Raphael Sgier for help with generating the data.
We acknowledge the support of the IT service of the Piz Daint computing cluster at the Swiss National Supercomputing Center (CSCS), as well as the Leonhard and Euler clusters at ETH Zurich.

\bibliographystyle{bmc-mathphys}
\bibliography{Refs}

\end{backmatter}

\final{\section*{Figures}}{}

\end{document}